\documentclass[reprint,pre,aps,superscriptaddress,longbibliography,amsmath]{revtex4-2}
\usepackage{graphicx,txfonts,dcolumn}
\usepackage[colorlinks,linkcolor=blue,citecolor=blue,anchorcolor=blue,urlcolor=blue]{hyperref}

\begin{document}

\title{Crossover Finite-Size Scaling Theory and Its Applications in Percolation}

\author{Ming Li}
\email{lim@hfut.edu.cn}
\affiliation{School of Physics, Hefei University of Technology, Hefei 230009, China}

\author{Sheng Fang}
\email{fs4008@mail.ustc.edu.cn}
\affiliation{School of Systems Science and Institute of Nonequilibrium Systems, Beijing Normal University, Beijing 100875, China}

\author{Jingfang Fan}
\email{jingfang@bnu.edu.cn}
\affiliation{School of Systems Science and Institute of Nonequilibrium Systems, Beijing Normal University, Beijing 100875, China}
\affiliation{Potsdam Institute for Climate Impact Research, Potsdam 14412, Germany}

\author{Youjin Deng}
\email{yjdeng@ustc.edu.cn}
\affiliation{Hefei National Research Center for Physical Sciences at the Microscale, University of Science and Technology of China, Hefei 230026, China}
\affiliation{Department of Modern Physics, University of Science and Technology of China, Hefei 230026, China}
\affiliation{Hefei National Laboratory, University of Science and Technology of China, Hefei 230088, China}

\date{\today}

\begin{abstract}
Finite-size scaling (FSS) for a critical phase transition ($t=0$) states that within a window of size $|t|\sim L^{-1/\nu}$, the scaling behavior of any observable $Q$ in a system of linear size $L$ asymptotically follows a scaling form as $Q(t,L)=L^{Y_Q}\tilde{Q}(tL^{1/\nu})$, where $\nu$ is the correlation-length exponent, $Y_Q$ is an FSS exponent and ${\tilde Q}(x)$ is a function of the scaled distance-to-criticality $x \equiv tL^{1/\nu}$. We systematically study the asymptotic scaling behavior of ${\tilde Q}(|x|\to\infty)$ for a broad variety of observables by requiring that the FSS and infinite-system critical behaviors match with each other in the crossover critical regime with $t \to 0$ and $|x|\to\infty$. This crossover FSS theory predicts that when the criticality is approached at a slower speed as $|t|\sim L^{-\lambda}$ with $\lambda <1/\nu$, the FSS becomes $\lambda$-dependent and the exponent can be derived. As applications, explosive percolation and high-dimensional percolation are considered. For the former, it is shown that the widely observed anomalous phenomena at the infinite-system criticality $t=0$ can be attributed to the mixing effects of the standard FSS behaviors around the pseudocritical point in an event-based ensemble. For the latter, FSS exponents are found to be different at the infinite-system critical and the pseudocritical point if free boundary conditions are used, and they are related to each other by using the crossover FSS theory. From these observations, the FSS of percolation systems falls into three classifications. Extensive simulations are carried out to affirm these predictions.
\end{abstract}

\maketitle

\section{Introduction}

Finite-size scaling (FSS) is a fundamental physical theory in the modern theory of phase transitions and critical phenomena~\cite{Privman1990,Ma2018}. It describes the size-dependent asymptotic scaling behaviors of finite systems to the thermodynamic limit in the neighborhood of a critical point. In the numerical study of phase transitions, the FSS theory provides a powerful and almost indispensable tool to locate the critical point, to determine the critical exponents, and to extract the associated universal properties~\cite{Binder1992}.

The core hypothesis of the FSS theory is as follows. Given an infinite system undergoing a continuous phase transition, it is known that, as the critical point is approached, the correlation length would diverge as $\xi \sim |t|^{-\nu}$, where $t$ characterizes the distance to the criticality and $\nu$ is the so-called correlation-length exponent. Then, for a finite system of linear length $L$, the FSS theory hypothesizes that the correlation length is smoothed out to be of order $\mathcal{O}(L)$. As a consequence, there exists a finite-size critical window $\mathcal{O}(L^{-1/\nu})$, and, within this window, an observable $Q$ would behave as
\begin{equation}
Q(t,L)=L^{Y_Q} \tilde{Q}(tL^{1/\nu}),   \label{eq-otl}
\end{equation}
where the FSS exponent $Y_Q$ is observable-dependent and $\tilde{Q}(x)$ is a function of the scaled distance-to-criticality $x\equiv tL^{1/\nu}$. Both $Y_Q$ and $\tilde{Q}(\cdot)$ are universal, i.e., independent of lattice types and other microscopic properties of the statistical physics models.

As an illustration of the FSS theory, we consider the bond percolation on a $d$-dimensional hypercubic lattice where each lattice bond is randomly occupied with some probability $P$ and any two sites connected through a chain of occupied bonds are said to be in the same cluster~\cite{Stauffer1994a}. As $P$ increases, the system undergoes a geometric percolation transition at $P_c$, from a subcritical phase composed of small clusters to a supercritical phase of a giant and spanning cluster. By denoting $t\equiv P-P_c$, Eq.~(\ref{eq-otl}) can be used to describe the FSS behaviors, near $t=0$, of a wide variety of geometric observables. A typical example is the second moment of cluster-size distribution $\chi \equiv L^{-d}\sum_{i\neq 1} C_i^2$, which sums over the sizes of all clusters $C_i$ except the largest one $C_1$ and plays a role as the magnetic susceptibility. The corresponding FSS exponent in Eq.~(\ref{eq-otl}) is $Y_{\chi} = 2d_f-d$, with $d_f$ the fractal dimension of critical clusters.

In some cases, besides the infinite-system percolation threshold $P_c$, it may be advantageous or practical to contemplate a pseudocritical point $P_L$ that depends on $L$. The pseudocritical point can be defined in various ways, such as the points where the susceptibility $\chi$ or the size $C_2$ of the second-largest cluster reaches their maximum values~\cite{Stauffer1994a}. Generally, it is conventionally assumed that pseudocritical points defined in different ways exhibit the same asymptotic behavior, characterized by $|P_L-P_c| \sim L^{-1/\nu}$. This implies that the FSS ansatz Eq.~(\ref{eq-otl}) can be equally applied in the vicinity of pseudocritical points, where the scaled distance-to-criticality $x\equiv tL^{1/\nu}$, concerning the infinite-system critical point and the pseudocritical point, is just shifted by an $L$-independent constant. Consequently, the same FSS can be extracted at the infinite-system critical and pseudocritical points. Indeed, this holds true in the majority of studies.

However, it was observed~\cite{Li2023,Li2024} that, in the vicinity of a dynamic pseudocritical point $\mathcal{P}_L$, the FSS behaviors of the so-called explosive percolation are rather clean and obey the standard FSS ansatz Eq.~(\ref{eq-otl}), while at $P_c$ and $P_L$, the finite-size behaviors are sophisticated and multiple fractal dimensions are observed. In addition, above the upper critical dimension $d_u=6$, the FSS behaviors for the standard percolation model are also very subtle and depend on the boundary condition. For instance, FSS exponents may take different values at $P_c$ and $P_L$ for free boundary conditions~\cite{Borgs2005,Borgs2005a,Heydenreich2006,Heydenreich2009}.

The goals of this work are twofold. Firstly, we formulate the crossover FSS theory by carefully examining critical behaviors of finite systems within a crossover critical window. This window is characterized by $(t\to0,x\to\infty)$ for FSS ansatz Eq.~(\ref{eq-otl}), representing a regime where the infinite-system critical point is gradually approached $(t\to0)$ while the rescaled distance-to-criticality diverges $(x\to\infty)$ with increasing system size. Secondly, based on the crossover FSS theory, we provide a quantitative explanation for the different FSS behaviors observed at $P_c$ and $\mathcal{P}_L$ in the explosive percolation and the high-dimensional percolation models. These two applications further suggest that FSS ansatz Eq.~(\ref{eq-otl}) is well applicable in the vicinity of the dynamic pseudocritical point $\mathcal{P}_L$, with the finite-size critical window spanning $\mathcal{O}(L^{-1/\nu})$ centered around $\mathcal{P}_L$.

The structure of the remaining sections is outlined as follows. In Sec.~\ref{sec-cfss}, we formulate the crossover FSS theory via a systematic examination of the crossover FSS behaviors for various physical observables in the framework of percolation phenomena. Applications on explosive percolation and high-dimensional percolation models are studied in Sec.~\ref{sec-app}, with a focus on the different FSS behaviors extracted at critical points of different definitions. Finally, Sec.~\ref{sec-dis} offers a summary to conclude the paper.

\section{Crossover finite-size scaling}  \label{sec-cfss}

In this section, we study the FSS behaviors in the crossover regime where the distance-to-criticality vanishes as $t\to0$, while the scaled distance-to-criticality diverges as $x\equiv tL^{1/\nu}\to\infty$. We formulate the crossover FSS theory by exploring the scaling form of the observable-dependent universal function $\tilde{Q}(x)$ in Eq.~(\ref{eq-otl}), as $x\to\infty$. Additionally, as an important corollary, we establish that when the distance-to-criticality vanishes as $|t| \sim L^{-\lambda} \gg L^{-1/\nu}$, the behaviors of observables for increasing $L$ can still be described by a power law with $\lambda$-dependent exponents, which we call finite-size \emph{lambda-scaling}.

\subsection{The thermodynamic scaling}

The critical behaviors of a continuous phase transition are characterized by the singularities of various physical quantities near the critical point $t=0$. These singularities are typically expressed in terms of power laws, which are governed by a set of critical exponents such as $\alpha$, $\beta$, $\gamma$, and others.

Generally, for spin models, like the $q$-state Potts model~\cite{Wu1982}, which reduces to the percolation model in the $q\to1$ limit, we can express the physical quantities in terms of the derivatives of the free-energy density $f(t,h)$ with respect to the thermal field $t$, and the magnetic field $h$. The critical exponents for the singularity of the specific heat $c$, the magnetization (order parameter) $m$, and the susceptibility $\chi$ with $t\to0$ are defined as follows:
\begin{align}
c(t) &\sim -\left.\frac{\partial^2 f}{\partial t^2}\right|_{h=0} \sim |t|^{-\alpha},  \label{eq-tc} \\
m(t) &\sim -\left.\frac{\partial f}{\partial h}\right|_{h=0} \sim t^{\,\beta},  ~~~~ t>0,  \label{eq-tm} \\
\chi(t) &\sim -\left.\frac{\partial^2 f}{\partial h^2}\right|_{h=0} \sim |t|^{-\gamma} \; ,
\end{align}
where $t>0$ is for the supercritical side, since the magnetization density is $0$ in the subcritical phase. In addition, the correlation length $\xi$ and the characteristic size $s_\xi$ of clusters, which can be well constructed for the percolation and the Fortuin-Kasterleyn representation of the Potts model, have the scalings
\begin{align}
\xi &\sim |t|^{-\nu},  \label{eq-txi}     \\
s_\xi &\sim |t|^{-1/\sigma}\sim \xi^{\,d_f},   \label{eq-tsxi}
\end{align}
where $d_f$ is the fractal dimension and $\sigma=1/\nu d_f$. Near the critical point $t=0$, the Fisher exponent $\tau$ can be introduced to characterize the power-law distribution of cluster sizes,
\begin{equation}
n(s) = s^{-\tau} \tilde{n}(s/s_\xi),   \label{eq-ns}
\end{equation}
which is defined as the number of clusters for a given size $s$ normalized by the system volume $V$. Here, the universal function $\tilde{n}(x)\to \mathcal{O}(1)$ for $x\to 0$, suggesting that $n(s)\sim s^{-\tau}$ at the critical point as $s_\xi\to\infty$. Note that $s_\xi$ can be regarded as the size of the largest cluster in the subcritical regime and of the second-largest cluster in the supercritical regime where the largest cluster is giant and percolating.

Indeed, even though the partition function remains constant for percolation, its critical behavior can still be understood by considering the $q\to1$ limit of the $q$-state Potts model~\cite{Wu1982}. In this limit, the specific heat vanishes, the order parameter is expressed as $m=C_1/V$, with $C_1$ the size of the largest cluster, and the susceptibility is $\chi=\sum_{i\neq 1}C_i^2$, where the sum $\sum_i$ runs over all the clusters with the largest one being excluded.

\subsection{Finite-size scaling}

According to the renormalization group theory, the FSS theory postulates that the free-energy density of a $d$-dimensional system with a finite side length $L$ can be expressed as
\begin{equation}
f(t,h,L)=f_r(t,h,L) + L^{-d}f_s(tL^{y_t},hL^{y_h},1),     \label{eq-fss}
\end{equation}
where $f_r$ is the regular part of the free-energy density, and $f_s$ is the singular part. The parameters $t$ and $h$ are respectively the thermal and magnetic scaling fields, and $y_t$ and $y_h$ are the corresponding thermal and magnetic rernormalization exponents, respectively.

Near criticality, the derivatives of the free-energy density Eq.~(\ref{eq-fss}) give the FSS,
\begin{align}
c(t,h,L) &\sim -\frac{\partial^2 f(t,h,L)}{\partial t^2}    \nonumber  \\
  &\sim L^{2y_t-d} f_s^{(2,0)} (tL^{y_t},hL^{y_h},1),   \\
m(t,h,L) &\sim -\frac{\partial f(t,h,L)}{\partial h}   \nonumber \\
  &\sim L^{y_h-d} f_s^{(0,1)} (tL^{y_t}, hL^{y_h},1),  \\
\chi(t,h,L) &\sim -\frac{\partial^2 f(t,h,L)}{\partial h^2}  \nonumber  \\
     &\sim L^{2y_h-d} f_s^{(0,2)} (tL^{y_t}, hL^{y_h},1).
\end{align}
In the absence of the external magnetic field ($h=0$), these functions can be simplified as
\begin{align}
c(t,L) &= L^{2y_t-d} \tilde{c}(tL^{y_t}),    \label{eq-fc}     \\
m(t,L) &= L^{y_h-d} \tilde{m}(tL^{y_t}),     \label{eq-fm}     \\
\chi(t,L) &= L^{2y_h-d} \tilde{\chi}(tL^{y_t}).    \label{eq-fchi}
\end{align}
Note that, the thermodynamic scaling of $m(t)$ as described in Eq.~(\ref{eq-tm}) is defined exclusively in the supercritical phase for infinite systems, since $m(t)$ is always zero in the subcritical phase. However, the finite-size behaviors of $m(t,L)$ outlined in Eq.~(\ref{eq-fm}) are applicable on both sides of criticality.

For finite systems, the divergence of the correlation length $\xi$ near criticality should be cut off by the side length $L$, that is $\xi\sim L$. Similar to Eqs.~(\ref{eq-fc})-(\ref{eq-fchi}), the FSS of the correlation length $\xi$ can be expressed as
\begin{equation}
\xi(t,L) = L\,\tilde{\xi}(tL^{y_t}).   \label{eq-fxi}
\end{equation}
According to Eq.~(\ref{eq-tsxi}), it also has
\begin{equation}
s_\xi(t,L) = L^{d_f}\tilde{s_\xi}(tL^{y_t}).   \label{eq-fsxi}
\end{equation}
It will be shown that $y_t=1/\nu$ and $y_h=d_f$. Thus, Eqs.~(\ref{eq-fc})-(\ref{eq-fsxi}) are indeed the specific forms of Eq.~(\ref{eq-otl}) for the corresponding physical quantities.

\subsection{Finite-size scaling in the crossover critical regime}

\begin{figure}
\centering
\includegraphics[width=\columnwidth]{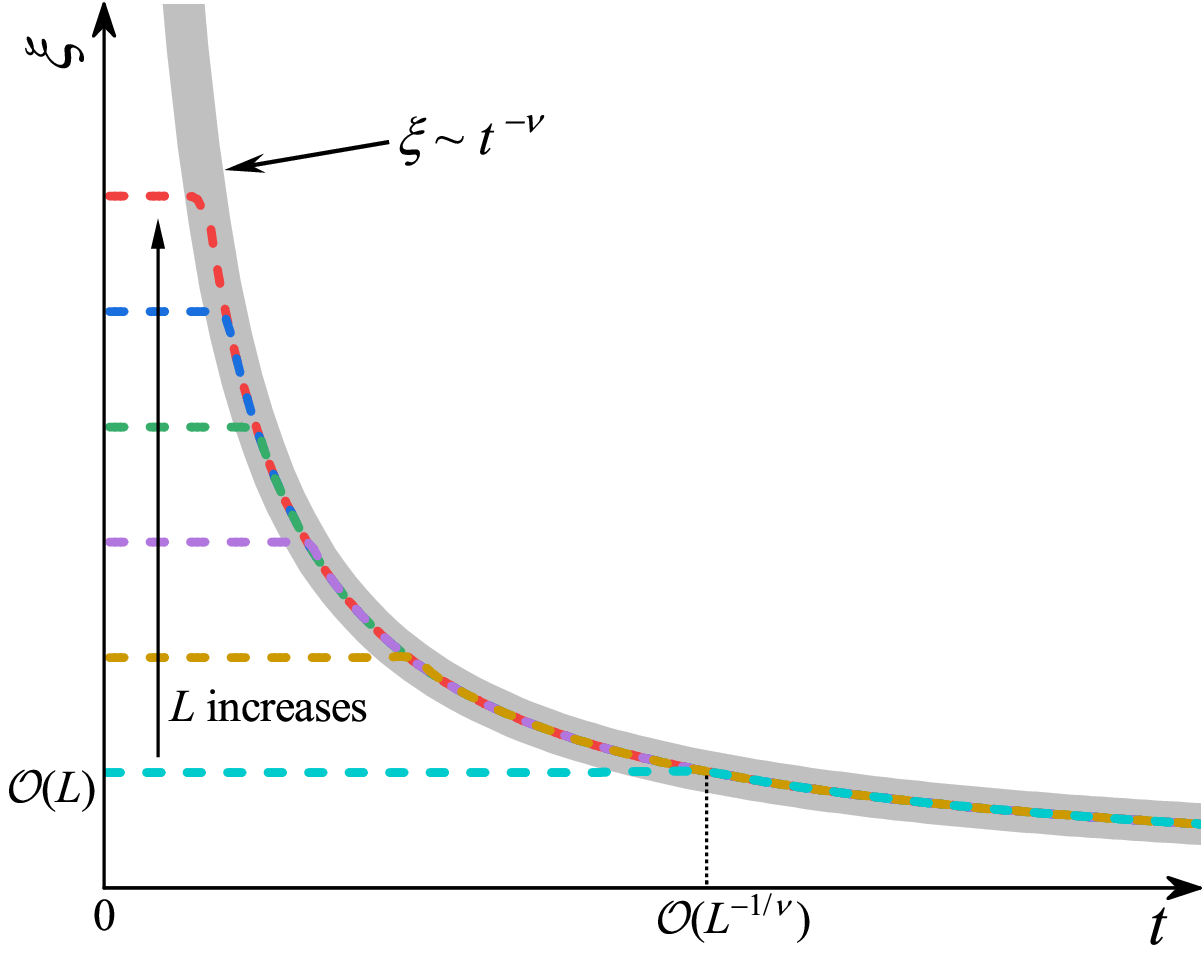}
\caption{(Color online) A sketch for the infinite-system and finite-size scaling behaviors of the correlation length $\xi$ in the supercritical phase. The grey thick line represents the thermodynamic scaling $\xi\sim |t|^{-\nu}$, while the dashed lines are for the finite-size behaviors $\xi(t,L)=L\,\tilde{\xi}(tL^{1/\nu})$ with different $L$. For finite systems, the divergency of $\xi$ for $|t|\to0$ saturates to the order of $\mathcal{O}(L)$, and there exists a turning point of $|t|\sim L^{-1/\nu}$, where the curve of $\xi(t,L)$ starts to separate from the curve of $\xi(t,\infty)\sim |t|^{-\nu}$.} \label{fg1}
\end{figure}

The FSS formulas Eqs.~(\ref{eq-fc})-(\ref{eq-fsxi}) naturally defines a finite-size critical window $|t|\sim \mathcal{O}(L^{-y_t})$ where the scaled distance-to-criticality $x \equiv tL^{y_t}$ takes a finite value for increasing $L$. The existence of such a finite-size critical window can be vividly illustrated by the growth of the correlation length $\xi$ as a system approaches the critical point. In Fig.~\ref{fg1}, we plot a sketch for the infinite-system and the finite-size behaviors of the correlation length $\xi$ near criticality. The thermodynamic scaling $\xi\sim |t|^{-\nu}$ suggests the divergence of $\xi$ as $|t|\to0$. For finite systems, $\xi(t,L)$ also grows as $|t|\to0$, eventually saturating to an order of $\mathcal{O}(L)$ near $t=0$. This behavior is captured by the FSS ansatz Eq.~(\ref{eq-fxi}). According to this ansatz, a correlation length $\xi$ of order $\mathcal{O}(L)$ can be observed not only precisely at $t=0$, but also at a finite scaled distance-to-criticality of $x\to\mathcal{O}(1)$. This defines the finite-size critical window of $\mathcal{O}(L^{-1/\nu})$, within which the FSS ansatz Eq.~(\ref{eq-otl}) is valid, and observable scales as $Q(t,L)\sim L^{Y_Q}$.

Beyond the finite-size critical window $t \gg L^{-y_t}$, the infinite-system critical behavior, $\xi \sim |t|^{-\nu}$, is expected to be observed as long as $\xi \gg 1$. For sufficiently large $L$, this defines a crossover critical regime with $L \gg \xi \gg 1$, which means $x=tL^{1/\nu} \gg 1$ and $t \to 0$. We assume that the FSS ansatz Eq.~(\ref{eq-fxi}) is asymptotically valid in this crossover critical regime $(t\to0,x\to\infty)$, where the FSS $\xi=L\tilde{\xi}(x)$ should be consistent with the infinite-system critical behavior $\xi\sim|t|^{-\nu}$. This implies that the universal function $\tilde{\xi}(x\to\infty)$ has to contain a term of $|t|^{-\nu}$, which further requires
\begin{equation}
\tilde{\xi}(x) \sim |x|^{\,-\nu},    ~~~|x|\to\infty.     \label{eq-xix}
\end{equation}
Substituting this formula into Eq.~(\ref{eq-fxi}), we get
\begin{align}
\xi(t,L) & = L\,\tilde{\xi}(x)        \nonumber \\
         & \sim L\, |x|^{-\nu}       \nonumber  \\
         & \sim |t|^{-\nu} L^{\,1-y_t\nu}.     \label{eq-tl}
\end{align}
By eliminating the $L$-dependent term in Eq.~(\ref{eq-tl}), it immediately recovers the thermodynamic scaling, as long as
\begin{equation}
y_t=\frac{1}{\nu}.        \label{eq-ytnu}
\end{equation}
This gives a relation between the critical exponents of the infinite-system critical behavior and the FSS behavior. So far, with Eq.~(\ref{eq-xix}) and the scaling relation $y_t=1/\nu$, the FSS Eq.~(\ref{eq-fxi}) reduces to the thermodynamic scaling Eq.~(\ref{eq-txi}) in the crossover critical regime $(t\to0,x\to\infty)$.

Similarly, in the crossover critical regime, to recover the thermodynamic scalings Eqs.~(\ref{eq-tc})-(\ref{eq-tsxi}) by the FSS Eqs.~(\ref{eq-fc})-(\ref{eq-fsxi}), it also requires the leading terms of the corresponding universal functions have the forms of
\begin{align}
\tilde{c}(x) &\sim |x|^{-\alpha},           \label{eq-cx}   \\
\tilde{m}^+(x) &\sim x^{\,\beta},  ~~~x>0,  \label{eq-mxp}   \\
\tilde{\chi}(x) &\sim |x|^{-\gamma},        \label{eq-chix}  \\
\tilde{s_\xi}(x) &\sim |x|^{-1/\sigma}.     \label{eq-sxix}
\end{align}
Since the thermodynamic scaling described in Eq.~(\ref{eq-tm}) is valid exclusively in the supercritical phase, the derived scaling of Eq.~(\ref{eq-mxp}) is applicable only for $t>0$, and the corresponding universal function is denoted as $\tilde{m}^+(x)$.

Substituting these forms of the universal functions into the corresponding FSS ansatz, the thermodynamic scaling will also be recovered, which further requires a zero exponent for the $L$-dependent term, as the discussion for Eq.~(\ref{eq-tl}). From this, more relations between the critical exponents of the thermodynamic scaling and the renormalization group are found,
\begin{align}
\alpha &= \frac{2y_t-d}{y_t},  \label{eq-ayt} \\
\beta &= \frac{d-y_h}{y_t},    \\
\gamma &= \frac{2y_h-d}{y_t},  \\
\sigma &= \frac{y_t}{d_f}.  \label{eq-sytdf}
\end{align}
With the scaling relations of the thermodynamic exponents, such as $\beta+\gamma=1/\sigma$, we can also find
\begin{equation}
d_f=y_h.           \label{eq-dfyh}
\end{equation}

Actually, the crossover critical behavior of the universal function $\tilde{m}^-(x)$ in the subcritical side is also available. In this regime, the order parameter $m(t,L)$ can be equivalently expressed as
\begin{equation}
m(t,L) \equiv \frac{C_1}{L^d} \sim \frac{\ln L}{L^d}  s_\xi(t,L),  \label{eq-msxi}
\end{equation}
where $C_1$ represents the largest-cluster size for the percolation model or the Fortuin-Kasterleyn representation of the Potts model. The presence of the logarithmic term $\ln L \sim C_1/s_\xi$ can be also understood within the framework of crossover FSS theory, as detailed in Sec.~\ref{sec-fd}. Then, comparing Eqs.~(\ref{eq-fm}), (\ref{eq-fsxi}), (\ref{eq-sxix}) and (\ref{eq-msxi}), we find that the leading term of the universal function $\tilde{m}^-(x)$ is given by
\begin{equation}
\tilde{m}^-(x) \sim |x|^{-1/\sigma}\ln L,  ~~~x<0.  \label{eq-mxn}
\end{equation}
It means that in the subcritical side of the crossover critical regime, the universal function $\tilde{m}(x)$ behaves similarly to $\tilde{s_\xi}(x)$, but with an additional logarithmic correction that scales with the system size. Substituting Eq.~(\ref{eq-mxn}) into Eq.~(\ref{eq-fm}), we obtain
\begin{align}
m(t,L) &\sim |x|^{-1/\sigma} L^{d_f-d} \ln L  \nonumber \\
       &\sim |t|^{-1/\sigma} L^{d_f-d-1/\sigma\nu} \ln L  \nonumber   \\
       &\sim |t|^{-1/\sigma} L^{-d} \ln L,
\end{align}
where $d_f=1/\sigma\nu$. This formula is consistent with the conventional understanding of the order parameter in the subcritical phase, indicating that $m(t,L)$ vanishes as $\sim \ln L/L^d$ due to the absence of long-range order.

\begin{figure}
\centering
\includegraphics[width=1.0\columnwidth]{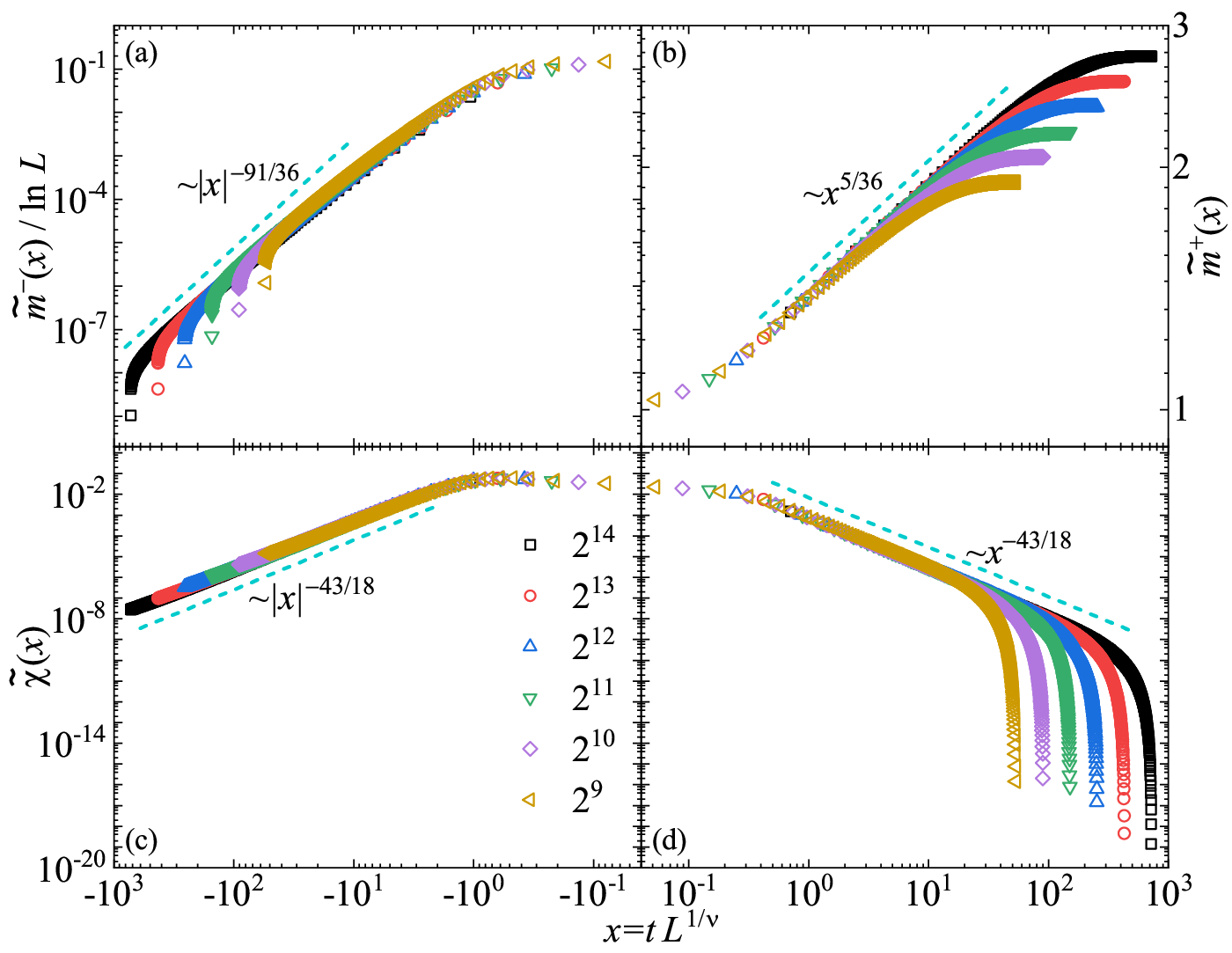}
\caption{(Color online) Crossover finite-size behaviors in terms of the scaled distance-to-criticality $x\equiv tL^{1/\nu}$ for the bond percolation on square lattices, with the correlation-length exponent $\nu=4/3$. (a) The rescaled universal function, $\tilde{m}^-(x)/\ln L \sim C_1(x)/L^{d_f}\ln L$, of the order parameter scales as $\sim x^{-1/\sigma}$ for $x\to-\infty$ and $L\to \infty$, with $\sigma=36/91$ indicated by the dashed line. (b) The universal function, $\tilde{m}^+(x)\sim C_1(x)/L^{d_f}$, of the order parameter scales as $\sim x^{\,\beta}$ for $x\to\infty$ and $L\to \infty$, with $\beta=5/36$ indicated by the dashed line. (c), (d) The universal function, $\tilde{\chi}(x)\sim \chi(x)/L^{2d_f-2}$, of the susceptibility scales as $\sim |x|^{-\gamma}$ for $|x|\to\infty$ and $L\to \infty$, with $\gamma=43/18$ indicated by the dashed lines.} \label{fg2}
\end{figure}

In summary, we have obtained the $x\to\infty$ asymptotic scaling behavior of the universal function $\tilde{Q}(x)$ for various observables $Q$. This crossover FSS theory, i.e., Eqs.~(\ref{eq-xix}), (\ref{eq-cx})-(\ref{eq-sxix}), and (\ref{eq-mxn}), delineates the $x$-dependent leading behavior of $\tilde{Q}(x)$. Together with these findings, we naturally establish the relationships between the critical exponents of the thermodynamic scaling and the renormalization group, as outlined in Eqs.~(\ref{eq-ytnu}) and (\ref{eq-ayt})-(\ref{eq-dfyh}).

As a numerical validation, we consider bond percolation on square lattices, for which the exact values of the renormalization exponents are known as $y_h=d_f=91/48$ and $y_t=1/\nu=3/4$. In Fig.~\ref{fg2} (b), representing the supercritical phase ($t>0$), the data of $\tilde{m}^+(x)$, calculated as $C_1(x)/L^{d_f}$, exhibit the expected scaling behavior $\sim x^{\,\beta}$ as per Eq.~(\ref{eq-mxp}), with $\beta=5/36$. Conversely, for $t<0$, as illustrated in Fig.~\ref{fg2} (a), the rescaled universal function $\tilde{m}^-(x)/\ln L$ asymptotically follows the scaling $\sim x^{-1/\sigma}$ with $\sigma=36/91$, in agreement with Eq.~(\ref{eq-mxn}). Regarding $\tilde{\chi}(x)$, shown in Figs.~\ref{fg2} (c) and (d), calculated as $\chi(x)/L^{2d_f-2}$, the scaling behavior $\sim |x|^{-\gamma}$ described by Eq.~(\ref{eq-chix}) is applicable to both sides of the crossover critical regime, with $\gamma=43/18$.

\subsection{Lambda-scaling}  \label{sec-lfss}

As an important corollary, the crossover FSS theory of Eqs.~(\ref{eq-xix}), (\ref{eq-cx})-(\ref{eq-sxix}), and (\ref{eq-mxn}) allows us to explore the FSS outside the standard finite-size critical window ($|t|\approx L^{-y_t}$). In particular, we study observables sampled at $|t|\sim L^{-\lambda}$ with $\lambda\leq y_t=1/\nu$.

Recalling Eq.~(\ref{eq-xix}) with $x=tL^{1/\nu} \sim L^{1/\nu-\lambda}$, the FSS of the correlation length sampled at $|t|\sim L^{-\lambda}$ can be found,
\begin{align}
\xi(t,L) & = L \, \tilde{\xi}(tL^{y_t})     \nonumber  \\
         & \sim L \, \left|tL^{y_t}\right|^{-\nu}     \nonumber  \\
         & \sim L^{\lambda\nu},      \label{eq-lxi}
\end{align}
where the relation $y_t=1/\nu$ is used. The FSS of other quantities at $|t|\sim L^{-\lambda}$ can be found similarly by calling Eqs.~(\ref{eq-cx})-(\ref{eq-sxix}), and (\ref{eq-mxn}),
\begin{align}
c(t,L) &\sim L^{\lambda\alpha},     \label{eq-lc}     \\
m^+(t,L) &\sim L^{-\lambda\beta}, ~~~ t>0,   \label{eq-lmp}   \\
m^-(t,L) &\sim L^{\lambda\nu d_f-d} \ln L, ~~~ t<0,   \label{eq-lmn}   \\
\chi(t,L) &\sim L^{\lambda\gamma},      \\
s_\xi(t,L) &\sim L^{\lambda/\sigma} \sim L^{\lambda\nu d_f}.   \label{eq-lsxi}
\end{align}
These scalings hold true for $\lambda<1/\nu$, i.e., in the crossover critical regime, and the derived FSS become $\lambda$-dependent, which we refer to as finite-size lambda-scaling.

\begin{figure}
\centering
\includegraphics[width=1.0\columnwidth]{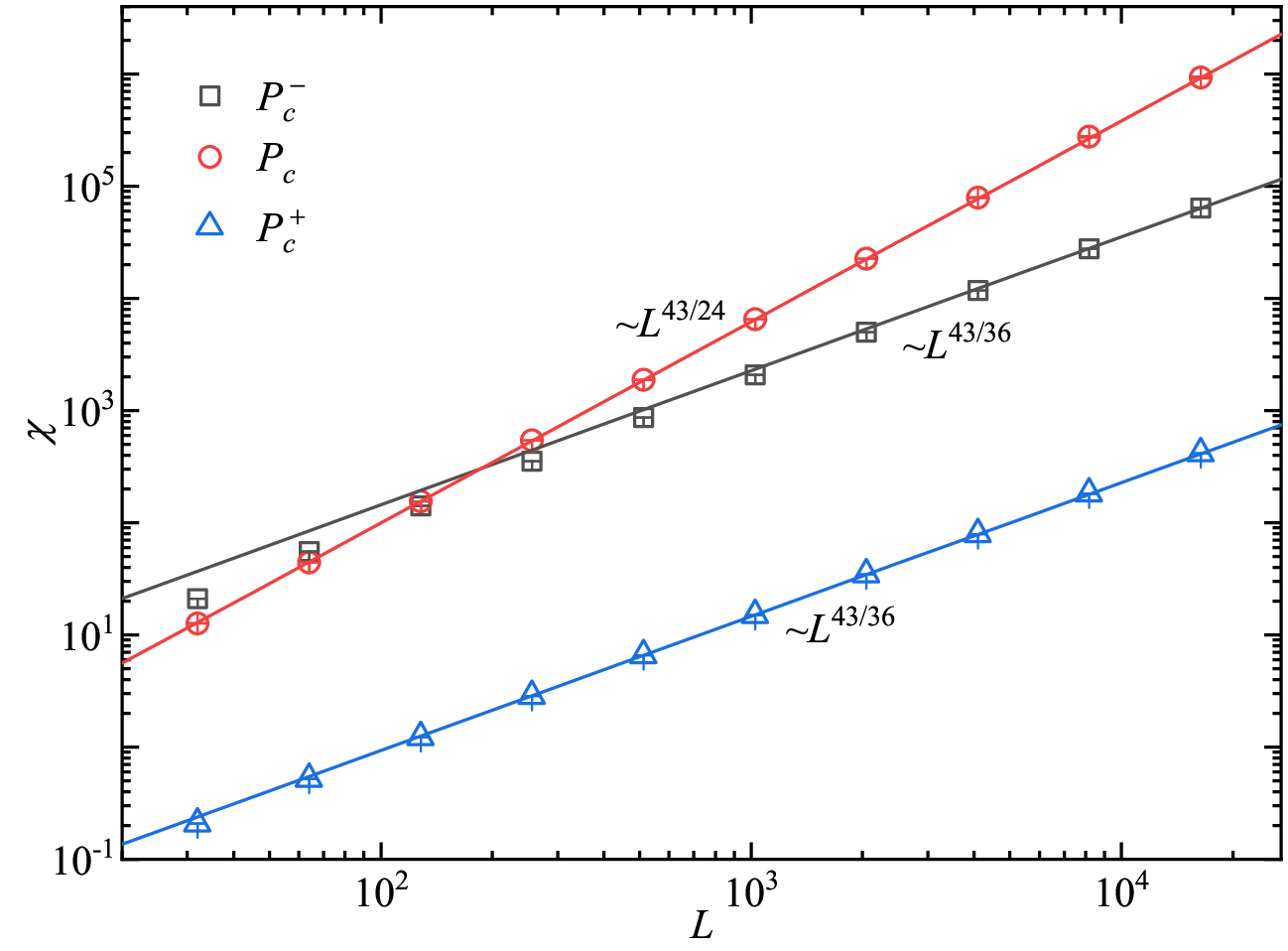}
\caption{(Color online) The FSS of the susceptibilities $\chi\sim L^{\gamma/\nu}$ sampled at $P_c=1/2$ and $\chi\sim L^{\lambda\gamma}$ sampled at $P_c^\pm\equiv P_c\pm aL^{-\lambda}$ with $\lambda=1/2$ for the bond percolation on square lattices. Without loss of generality, the parameter $a$ is set to be $0.5$. All these scalings are consistent with the finite-size lambda-scaling prediction indicated by the solid lines. At $P_c$, the standard FSS can be observed with exponent $\gamma/\nu=43/24$. For $P_c^\pm$, the lambda-scaling with exponent $\lambda\gamma=43/36$ is observed.} \label{fg3}
\end{figure}

To demonstrate the finite-size lambda-scaling, we consider bond percolation on square lattices, and plot in Fig.~\ref{fg3} the results of the susceptibility $\chi$ sampled at $P_c=1/2$ and $P^\pm_c\equiv P_c\pm aL^{-\lambda}$ as a function of the side length $L$, where $a$ is an $L$-independent coefficient. Without loss of generality, we set the parameter $a$ is set to be $0.5$ in the simulation, and $\lambda=1/2$ is chosen to make the system out of the standard finite-size critical window. At the infinite-system critical point $P_c=1/2$, the susceptibility $\chi$ shows the scaling $\chi\sim L^{\gamma/\nu}$ with the standard critical exponents of $2$-dimensional universality $\gamma/\nu=43/24$ from $\gamma=43/18$ and $\nu=4/3$. At $P^\pm_c$, the system is outside the finite-size critical window, and Fig.~\ref{fg3} clearly shows a different FSS from the one extracted at $P_c$, corresponding to the finite-size lambda-scaling of $\chi\sim L^{\lambda\gamma}$ with $\lambda\gamma = 43/36$.

To analyze the data, we fit the data of $\chi$ to the scaling ansatz
\begin{equation}
Q(L) = L^{Y_Q} (a_0+a_1L^{-\omega_1}+a_2L^{-\omega_2}),   \label{eq-fit1}
\end{equation}
where $L^{Y_Q}$ describes the leading behavior, and $a_iL^{-\omega_i}~(i=1,2)$ are for the finite-size corrections. In the fit, we impose a lower cutoff $L \geq L_{\text{min}}$ on the data points and systematically study the fits (the goodness and the stability) by increasing $L_{\text{min}}$. Generally, the preferred fit for any given ansatz corresponds to the smallest $L_{\text{min}}$ for which the residual error chi-square is less than one unit per degree of freedom and subsequent increases in $L_{\text{min}}$ do not cause the chi-square value to drop by vastly more than one unit per degree of freedom. This method is applied to all the fit in this paper. With one finite-size correction term ($a_2=0$), we obtain stable fit results of lambda-scaling $\chi\sim L^{\lambda\gamma}$ in Fig.~\ref{fg3}, i.e., $\lambda\gamma=1.13(6)$ and $1.196(2)$ for $P^-_c$ and $P^+_c$, respectively, which agree well with the theoretical value $\lambda\gamma = 43/36\approx1.194$ predicted by the crossover FSS theory.

\subsection{Lambda-scaling exponent of the largest cluster}  \label{sec-fd}

With the finite-size lambda-scaling, we further study the fractal dimension of the largest cluster. For the percolation model and the Fortuin-Kasterleyn representation of the Potts model, the order parameter $m(t,L)$ can be defined as the occupation fraction of the largest cluster over the whole lattice, i.e., $m(t,L)\equiv C_1/L^d$, thus, the FSS behavior of $C_1$ can be expressed as
\begin{equation}
C_1=L^{d}m(t,L).          \label{eq-c1fss}
\end{equation}
According to Eqs.~(\ref{eq-lmp}) and (\ref{eq-lmn}), we have
\begin{align}
C_1^+ &= L^{d-(d-d_f)\lambda\nu},~~~t>0,  \label{eq-c1u}        \\
C_1^- &= L^{\lambda\nu d_f} \ln L,~~~~~t<0.  \label{eq-c1l}
\end{align}
Therefore, the lambda-scaling fractal dimension $d_\lambda$ of the largest cluster in the crossover critical regime is obtained,
\begin{align}
d_\lambda^+ & = d-(d-d_f)\lambda\nu,~~~t>0,  \label{eq-dlup} \\
d_\lambda^- & = \lambda\nu d_f,~~~~~~~~~~~~~~~~~~t<0.  \label{eq-dllow}
\end{align}

\begin{figure}
\centering
\includegraphics[width=1.0\columnwidth]{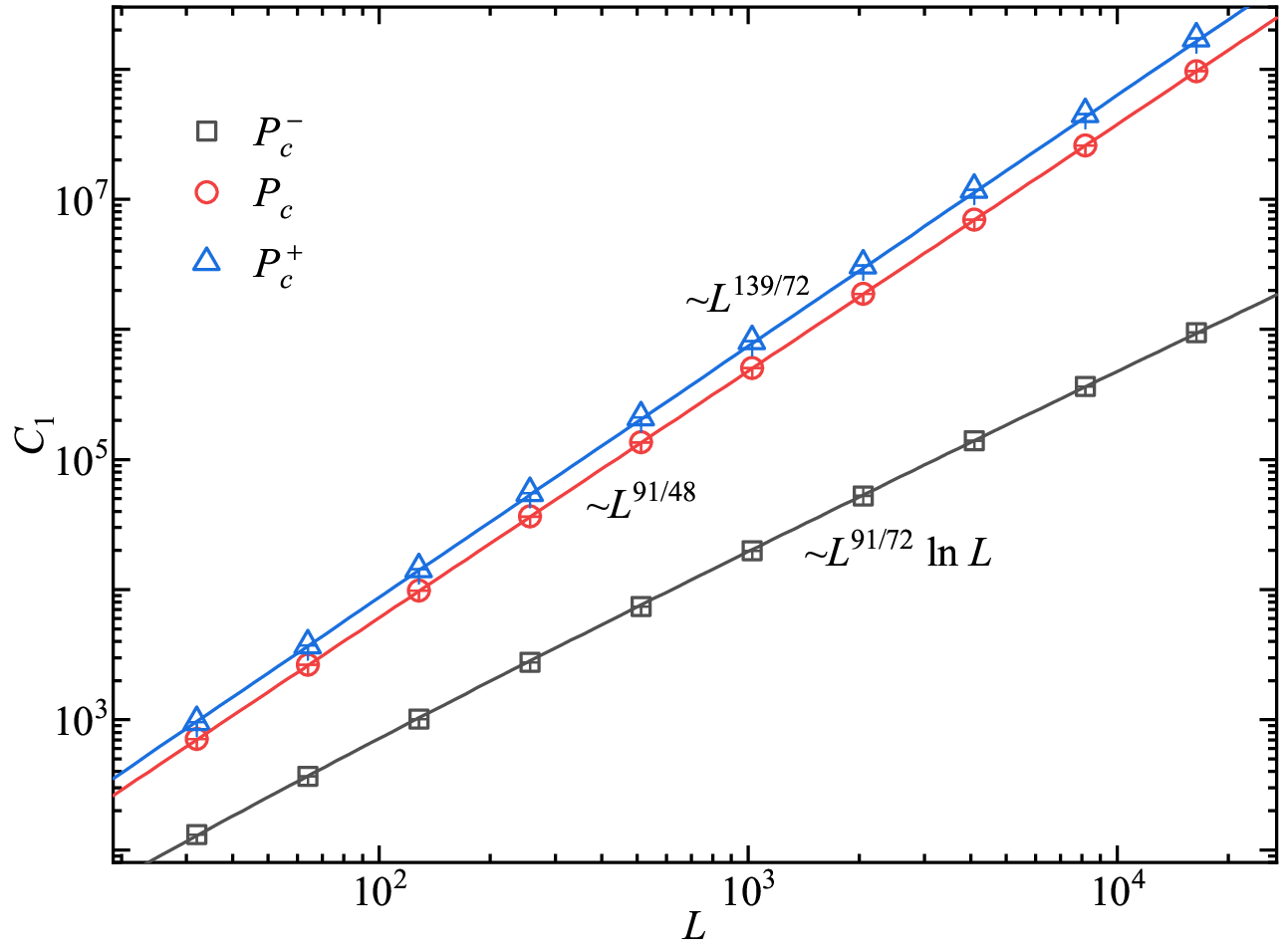}
\caption{(Color online) The FSS of the size $C_1$ of the largest cluster sampled at $P_c=1/2$ and $P_c^\pm\equiv P_c\pm aL^{-\lambda}$ with $\lambda=1/2$ for the bond percolation on square lattices. Without loss of generality, the parameter $a$ is set to be $0.5$. According to Eqs.~(\ref{eq-dlup}) and (\ref{eq-dllow}), the fractal dimensions of the largest-cluster size $C_1^\pm$ can be derived, $d_\lambda^+=139/72$ and $d_\lambda^-=91/72$, from the fractal dimension $d_f=91/48$ of $2$-dimensional universality, which is observed at $P_c$. By fitting the data sampled at $P^+_c$ and $P^-_c$ to the scaling ansatz Eqs.~(\ref{eq-fit1}) and (\ref{eq-fit2}) with $a_2=0$, respectively, we find the lambda-scaling $d_\lambda^+=1.9305(1)$ and $d_\lambda^-=1.26(1)$, which are consistent with the theoretical prediction indicated by the solid lines. Note that at $P_c^-$, the FSS of $C_1^-$ has a multiplicative logarithmic correction.} \label{fg4}
\end{figure}

The FSS of $C_1^\pm$ sampled at $P^\pm_c\equiv P_c\pm aL^{-\lambda}$, which are asymmetric at the supercritical and the subcritical side of $P_c$, is plotted in Fig.~\ref{fg4} for the bond percolation on square lattices. With $\lambda=1/2$, the system is outside the finite-size scaling window, and the resulting FSS are notably different from the standard one at $P_c$. By fitting the data of $C_1^+$ to the scaling ansatz Eq.~(\ref{eq-fit1}) with $a_2=0$, we find the lambda-scaling $d_\lambda^+=1.9305(1)$. This result is in good agreement with the theoretical value predicted by Eqs.~(\ref{eq-dlup}), i.e., $139/72\approx1.9306$. For $C_1^-$, we fit the data to the scaling ansatz
\begin{equation}
Q(L) = L^{Y_Q}(\ln L +b)(a_0+a_1L^{-\omega_1}+a_2L^{-\omega_2}). \label{eq-fit2}
\end{equation}
With $a_2=0$, we find $d_\lambda^-=1.26(1)$, which is also consistent with the lambda-scaling of Eq.~(\ref{eq-dllow}), i.e., $91/72\approx1.2639$.

The observed logarithmic correction to the finite-size scaling of $C_1^-$ in Fig.~\ref{fg4} can be attributed to the non-uniqueness of clusters of size $\mathcal{O}(L^{d_\lambda})$ in the subcritical side of the crossover critical regime. From Eq.~(\ref{eq-ns}), in the crossover critical regime $(t\to0,x\to\infty)$ with $|t|\sim L^{-\lambda}$, it is expected that the cluster number density takes the scaling form
\begin{equation}
n(s,L) = s^{-\tau}\tilde{n}(s/L^{d_\lambda}),        \label{eq-nslb}
\end{equation}
where $L^{d_\lambda}$ refers to the characteristic size $s_\xi(\lambda)<L^{d_f}$. Note that the universal function $\tilde{n}(x\equiv s/L^{d_\lambda})$ decays quickly as $x\gg 1$, and the Fisher exponent remains to be $\tau=1+d/d_f$, and is larger than $2$. Thus, by integrating Eq.~(\ref{eq-nslb}) from $L^{d_\lambda}$ to $L^d$, we can estimate the number $N_c$ of clusters with characteristic size $s_\xi(\lambda)\sim L^{d_\lambda}$ as
\begin{align}
N_c(\lambda) &= L^d \int_{L^{d_\lambda}}^{L^d} s^{-\tau} \tilde{n}(s/L^{d_\lambda}) ds  \nonumber \\
             &\sim L^{d(1-d_\lambda/d_f)}     \\
             &\sim L^{d(1-\lambda\nu)},    \label{eq-ncl}
\end{align}
where $\tau=1+d/d_f$ and $d_\lambda=\lambda\nu d_f$ are used. The divergence of $N_c(\lambda)$ is immediately evident for $\lambda<1/\nu$ outside the finite-size critical window. Within the finite-size critical window, however, one has $d_\lambda=d_f$ and $N_c(\lambda)\sim\mathcal{O}(1)$, indicating a finite number of clusters of size $\mathcal{O}(L^{d_f})$.

From the extreme-value theory, the largest-cluster size $C_1$ among a divergence number $N_c(\lambda)$ of clusters of size $\mathcal{O}(L^{d_\lambda}) $ can be estimated by
\begin{equation}
L^d \int_{C_1}^{L^d} s^{-\tau} \tilde{n}(s/L^{d_\lambda}) ds \sim \mathcal{O}(1).   \label{eq-c1int}
\end{equation}
Near the upper bound of the integral, $L^d \gg L^{d_\lambda}$, the universal function $\tilde{n}(s/L^{d_\lambda})$ vanishes. The lower bound of the integral, $C_1$, is slightly larger than the average size of the characteristic clusters, which generates a fast-decaying universal function. Let us assume that the universal function $\tilde{n}(x\equiv s/L^{d_\lambda})$ has an exponential decay for $x>1$~\cite{Stauffer1994a}. As an estimate for large $L$, the integral Eq.~(\ref{eq-c1int}) can be simplified as
\begin{equation}
L^d  C_1^{1-\tau} e^{-C_1/L^{d_\lambda}} \sim \mathcal{O}(1).
\end{equation}
By solving this equation, we can immediately recover the leading behavior of $C_1$ as given in Eq.~(\ref{eq-c1l}), namely, $C_1 \sim L^{\lambda\nu d_f} \ln L$. This elucidates the origin of the logarithmic term in Eq.~(\ref{eq-msxi}), governing the behavior of $m(t,L)$ in the subcritical side of the crossover critical regime.

Additionally, it is noteworthy that the scaling described by Eq.~(\ref{eq-c1l}) must also apply to the second largest cluster in the supercritical crossover regime. This is because the starting point of above discussions, Eq.~(\ref{eq-nslb}), remains valid for $t>0$, with the largest cluster excluded. In essence, this implies that the two largest clusters in the supercritical regime possess different length scales, as described by Eqs.~(\ref{eq-c1u}) and (\ref{eq-c1l}), respectively.

The decay of the universal function $\tilde{n}(s/s_\xi)$ in Eq.~(\ref{eq-nslb}) for $s>s_\xi$ could generally follow a stretched exponential function, depending on the size distribution of clusters of size $s_\xi$. Consequently, the specific form of the resulting logarithmic term, as that in Eqs.~(\ref{eq-msxi}) and (\ref{eq-c1l}), might be more intricate, potentially assuming a general form of $(\ln L/L_0)^\kappa$ with model-dependent parameters $L_0$ and $\kappa$.

\section{Applications}  \label{sec-app}

In this section, we apply the crossover FSS theory to explosive percolation on complete graphs and high-dimensional percolation on hypercubic lattices with free boundary conditions. By meticulously examining the FSS behaviors at the pseudocritical and infinite-system critical points, a classification of percolation systems is also obtained.

\subsection{Explosive percolation}

Explosive percolation, typically defined on a complete graph, begins with an empty graph devoid of any bonds. At each time step, two candidate bonds are randomly selected from all un-inserted ones. Subsequently, the size product of the two clusters branching from the ends of a bond is computed, and the bond with the smaller size product is inserted. The system eventually percolates as bonds are continually inserted, characterized by a sudden emergence of a large and dense cluster, with a plethora of anomalous critical phenomena, sometimes even leading to the misconception of a discontinuous transition~\cite{Achlioptas2009,Friedman2009,Ziff2009,Radicchi2010,Grassberger2011,Tian2012,DSouza2015,DSouza2019}. Recently, employing an event-based ensemble~\cite{Li2023,Li2024}, it has been revealed that explosive percolation can indeed be captured by the standard FSS theory around the dynamic pseudocritical point.

To pinpoint the percolation threshold of explosive percolation, we monitor the one-step incremental size, denoted as $\Delta(\mathcal{B}) \equiv \mathcal{C}_1(\mathcal{B}+1)-\mathcal{C}_1(\mathcal{B})$~\cite{Nagler2011,Manna2011,Fan2020}, where $\mathcal{C}_1(\mathcal{B})$ is the size of the largest cluster at time step $\mathcal{B}$. As time step progresses, $\Delta(\mathcal{B})$ generally increases during the subcritical phase, reaching its peak at a certain time step $\mathcal{B}_{\text{max}}$, before decreasing as the system enters into the supercritical phase. Consequently, the bond density $\mathcal{P}_{L}$, calculated as $\mathcal{B}_{\text{max}}/V$ (with $V$ representing the total number of sites in the system), serves as the pseudocritical point. Given that $\mathcal{P}_{L}$ varies across realizations, it is referred to as the dynamic pseudocritical point. Furthermore, the mean of dynamic pseudocritical points $P_L\equiv\langle\mathcal{P}_L\rangle$ corresponds to a static pseudocritical point of traditional definition, approaching the infinite-system critical point $P_c$ as the system size increases.

\begin{figure}
\centering
\includegraphics[width=\columnwidth]{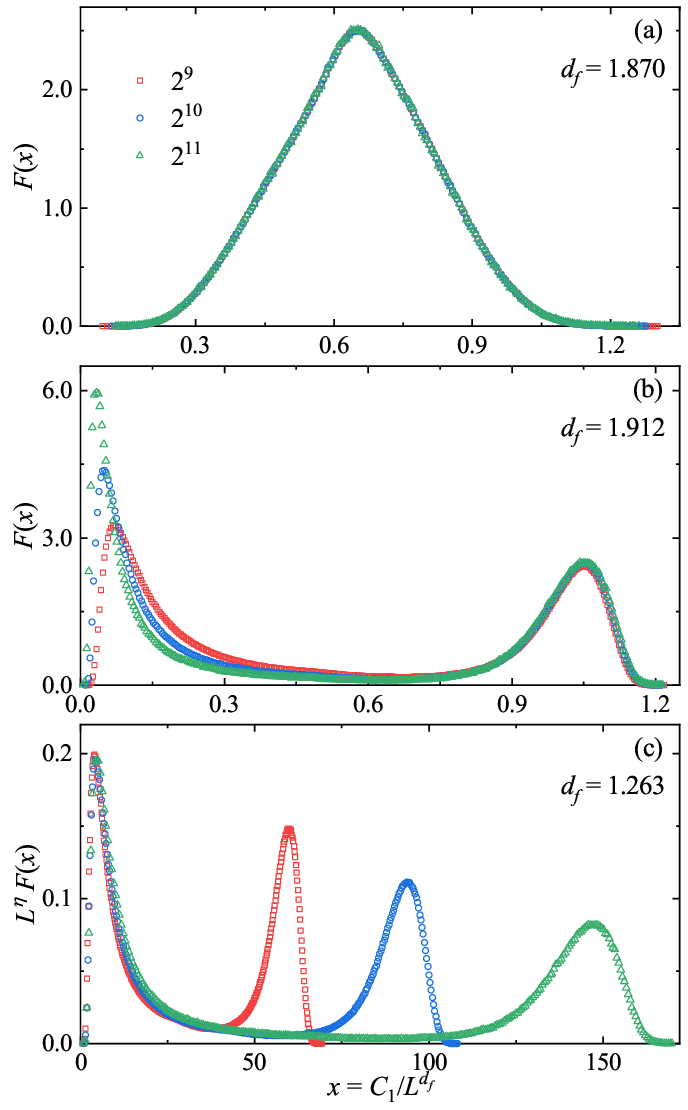}
\caption{(Color online) The distribution $F(x)$ of the critical size $C_1$ of the largest cluster for explosive percolation on complete graphs of different side length $L$. For the consistency and convenience of the definition of FSS, the side length used here is defined as $L\equiv\sqrt{V}$, where $V$ is the total number of sites in the complete graph. (a) At $\mathcal{P}_L$, a good data collapse can be achieved by defining $x\equiv C_1/L^{d_f}$ with $d_f=1.87$. (b), (c) At $P_c=0.8884491$~\cite{Li2023,Li2024}, a bimodal distribution $F(x)$ can observed, and two distinguishable fractal dimensions $d_f=1.912$ and $d_f=1.263$ are needed to roughly collapse the data near the two peaks, respectively. Note that neither of the two fractal dimensions takes the correct value $d_f=1.870$ at the dynamic pseudocritical point. For the left peak, a rescaled exponent $\eta=0.2$ is also used to show a better data collapse.} \label{fg5}
\end{figure}

In Fig.~\ref{fg5}, we illustrate the probability distribution function $F(x)$ of the largest-cluster size $\mathcal{C}_1$ at both $\mathcal{P}_L$ and $P_c=0.8884491$~\cite{Li2023,Li2024} for different system sizes $L$. Since the complete graph has no inherent side length for the definition of FSS, we artificially create a side length of $L\equiv\sqrt{V}$, by embedding all sites into a square lattice of volume $V$. At $\mathcal{P}_L$, the distributions $F(x)$ for various $L$ exhibit excellent collapse onto each other when defining $x\equiv C_1/L^{d_f}$ with a unique fractal dimension $d_f=1.87$, which corresponds to the FSS of $C_1\sim L^{d_f}$ extracted at $\mathcal{P}_L$~\cite{Li2023}. This indicates the effective applicability of the standard FSS theory to the dynamic pseudocritical point $\mathcal{P}_L$. However, as depicted in Figs.~\ref{fg5} (b) and (c), the $F(x)$ distribution at $P_c$ displays a bimodal pattern, requiring two distinct fractal dimensions to achieve a rough data collapse around the two peaks, respectively. Such a $\mathcal{C}_1$ distribution is evidently not accounted for by the standard FSS theory and is often categorized as anomalous FSS behaviors~\cite{Grassberger2011}.

We further study the asymptotic behavior of $\mathcal{P}_L$ for large $L$, by examining the FSS behaviors of the mean deviation of $\mathcal{P}_L$ from $P_c$ and its variance~\cite{Fan2020},
\begin{align}
\delta(\mathcal{P}_L) &\equiv \left|\langle \mathcal{P}_L\rangle-P_c\right| \sim L^{-X_\text{m}},  \label{eq-plm}          \\
\text{Var}(\mathcal{P}_L) &\equiv \langle \mathcal{P}_L^2\rangle-\langle \mathcal{P}_L\rangle^2 \sim L^{-2X_{\text{v}}}, \label{eq-plsd}
\end{align}
where two exponents $X_\text{m}$ and $X_{\text{v}}$ are introduced for the mean deviation and variance, respectively. According to the theory of critical phenomena, the correlation length is the only length scale, and it is expected that both the mean deviation of $\mathcal{P}_L$ from $P_c$ and its fluctuation are governed by the correlation-length exponent $\nu$, i.e., $X_\text{m}=X_\text{v}=1/\nu$. Indeed, this holds true in most cases, such as the bond percolation below the upper critical dimension $d_u=6$.

\begin{figure}
\centering
\includegraphics[width=\columnwidth]{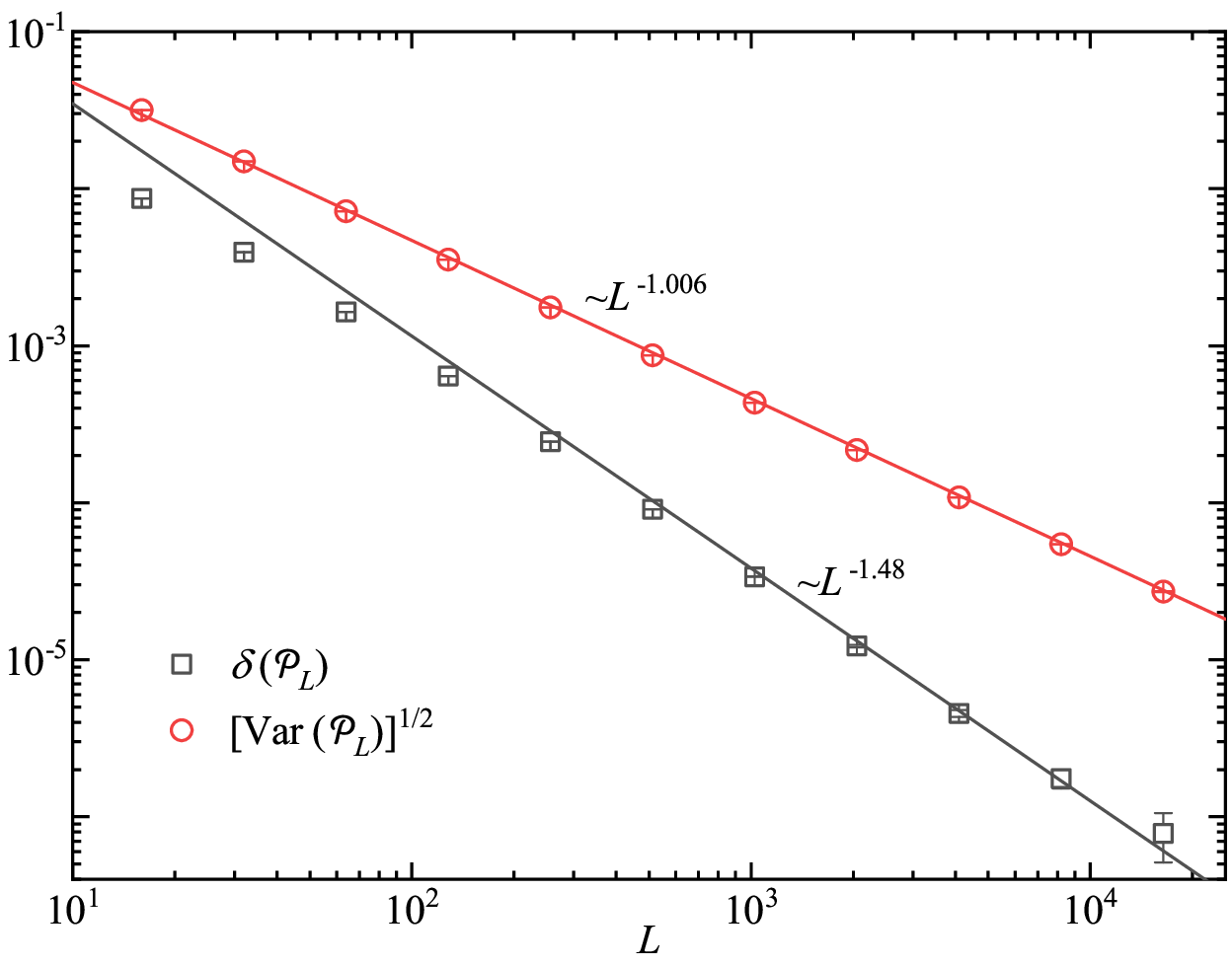}
\caption{(Color online) The finite-size behaviors of the dynamic pseudocritical point $\mathcal{P}_L$ for explosive percolation on complete graphs. Here, the FFS of the mean deviation $\delta(\mathcal{P}_L)\sim L^{-X_\text{m}}$ of $\mathcal{P}_L$ from $P_c=0.8884491$~\cite{Li2023,Li2024}, and its fluctuation $\sqrt{\text{Var}(\mathcal{P}_L)}\sim L^{-X_\text{v}}$ are presented in the log-log plot, where the lines show the fit results of $X_\text{m}=1.480(4)$ and $X_\text{v}=1.006(6)$, obtained by fitting the data to the scaling ansatz Eq.~(\ref{eq-fit1}) with $a_2=0$. For the consistency and convenience of the definition of the FSS, the side length used here is defined as $L=\sqrt{V}$, where $V$ is the total number of sites in the complete graph, and the occupied probability of bonds $P$ is defined as the ratio between the number of occupied bonds and the system volume $V=L^2$.} \label{fg6}
\end{figure}

In Fig.~\ref{fg6}, we present the finite-size behaviors of the dynamic pseudocritical point $\mathcal{P}_L$ for the explosive percolation on the complete graph. In contrast to one's expectation, it can be clearly seen that in the log-log plot, $\delta(\mathcal{P}_L)$ and $\sqrt{\text{Var}(\mathcal{P}_L)}$ vanish with different slopes, clearly illustrating the different exponents $X_\text{m}$ and $X_\text{v}$. By fitting the data to the scaling ansatz Eq.~(\ref{eq-fit1}) with $a_2=0$, we determine the exponent $X_\text{m}=1.480(4)$, notably larger than $X_\text{v}=1.006(6)$.

\begin{figure}
\centering
\includegraphics[width=1.0\columnwidth]{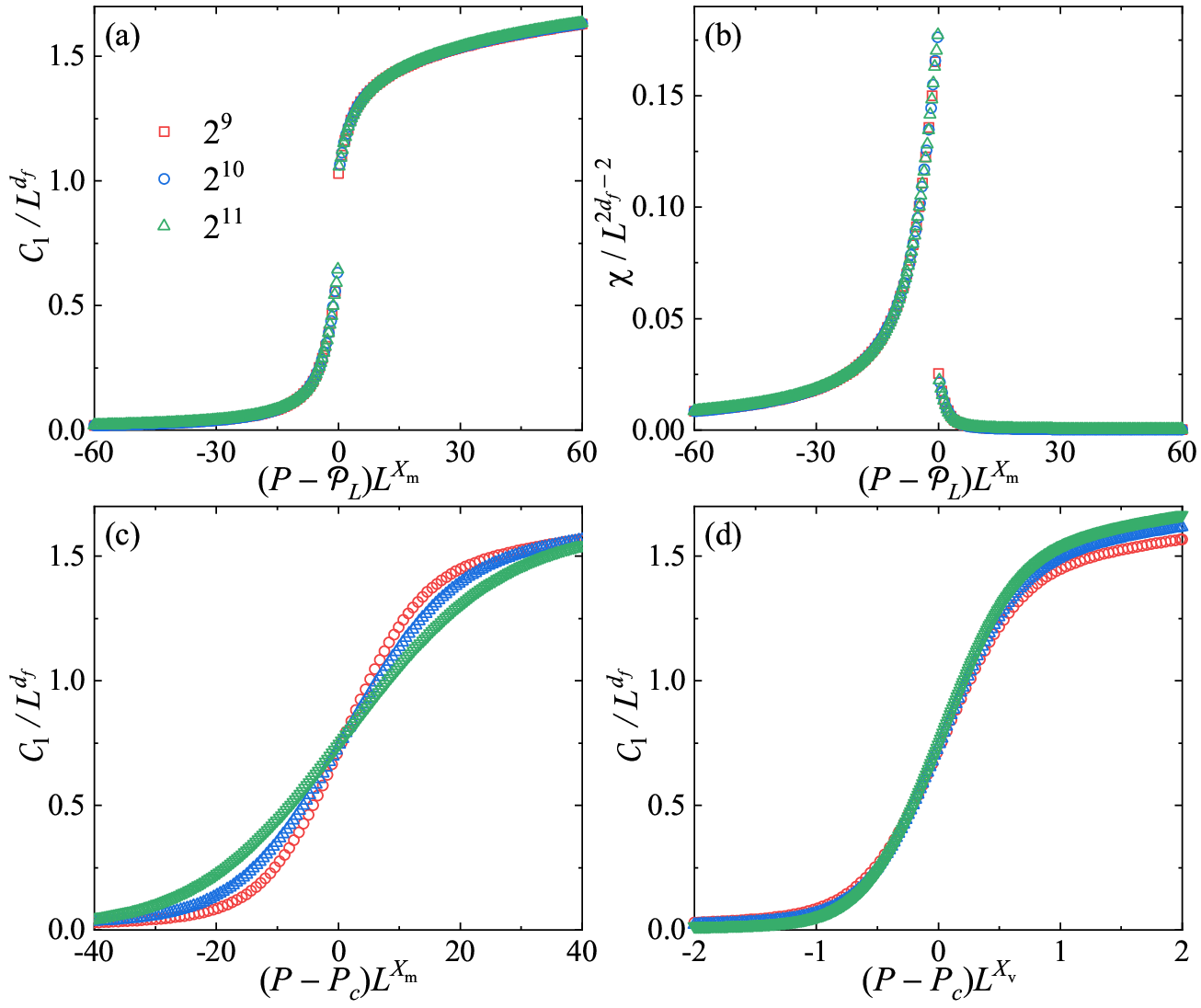}
\caption{(Color online) The data collapse near the dynamic pseudocritical point $\mathcal{P}_L$ and the infinite-system critical point $P_c$ for explosive percolation on complete graphs. The scatters of different colors and shapes represent systems of different side lengths $L$, which is defined as $L\equiv\sqrt{V}$, with $V$ the total number of sites. (a), (b) The plots of $C_1/L^{d_f}$ and $\chi/L^{2d_f-2}$ versus $(P-\mathcal{P}_L)L^{1/\nu}$ with $1/\nu=X_\text{m}=1.48$ and $d_f=1.8697$ for different $L$. The discontinuity at $\mathcal{P}_L$ arises from its event-based definition. (c), (d) The plots of $C_1/L^{d_f}$ versus $(P-P_c)L^{1/\nu}$ with $1/\nu=X_\text{m}=1.48$ and $1/\nu=X_\text{v}=1.006$ for different $L$. The correct fractal dimension $d_f=1.8697$ is used here. The good data collapse near $\mathcal{P}_L$ suggests that the FSS of explosive percolation is well-defined around $\mathcal{P}_L$, with correlation-length exponent $\nu=1/X_\text{m}$, rather than $P_c$.} \label{fg7}
\end{figure}

Furthermore, we examine the FSS ansatz Eq.~(\ref{eq-otl}) for both $1/\nu=X_\text{m}$ and $1/\nu=X_{\text{v}}$. In Figs.~\ref{fg7} (a) and (b), we present the renormalized largest-cluster size $C_1/L^{d_f}$ and the susceptibility $\chi/L^{2d_f-2}$ versus $x\equiv tL^{X_\text{m}}$, where $t\equiv P-\mathcal{P}_L$. It is clear that the data for different $L$ can collapse onto each other over a wide range of $x$, supporting the notion that the system near $\mathcal{P}_L$ adheres to the standard FSS. From this, we further argue that the correlation-length exponent is $\nu=1/X_\text{m}$ for explosive percolation. In contrast, the data collapse cannot be achieved around $P_c$, regardless of whether $X_\text{m}$ or $X_{\text{v}}$ is chosen as the correlation-length exponent $1/\nu$, as shown in Figs.~\ref{fg7} (c) and (d).

The good collapse shown in Figs.~\ref{fg7} (a) and (b) defines a finite-size critical window of size $O(L^{-X_\text{m}})$, round $\mathcal{P}_L$. The smaller exponent $X_\text{v}$ compared to $X_\text{m}$ implies a larger fluctuation window for $\mathcal{P}_L$. Consequently, for some realizations, $P_c$ can fall outside the finite-size critical window $\mathcal{O}(L^{-X_\text{m}})$, which is now centered on $\mathcal{P}_L$. In other words, some of the FSS behaviors extracted at $P_c$ correspond to those exceeding the finite-size critical window around $P_L$, which can be describe by the crossover FSS theory. More precisely speaking, in accordance with the lambda-scaling of Eqs.~(\ref{eq-dlup}) and (\ref{eq-dllow}), we expect to have $\lambda=X_\text{v}=1$ and $1/\nu=X_\text{m}=1.48$. By substituting the fractal dimension $d_f=1.8697$ extracted at $\mathcal{P}_L$~\cite{Li2023}, we can derive the two extremes of the fractal dimensions extracted at $P_c$: $d_\lambda^+=2-(2-d_f)\lambda\nu=1.912$, and $d_\lambda^-=\lambda\nu d_f=1.263$. This explains the bimodal distribution of the size of the largest cluster in Fig.~\ref{fg5}, where $d_\lambda^\pm$ precisely corresponds to the fractal dimensions for samples near the two peaks. In fact, the largest cluster extracted at $P_c$ comprises samples with continuously varying fractal dimensions lying between $d_\lambda^+=1.912$ and $d_\lambda^-=1.263$, representing the samples in the large fluctuation window $\mathcal{O}(L^{-X_\text{v}})$. Hence, the FSS behaviors of physical observables sampled at $P_c$ reflect the stochastic mixing effects of those within a wide neighborhood range of $\mathcal{P}_L$. In terms of $x\equiv (P-\mathcal{P}_L)L^{1/\nu}$, this range encompasses the finite-size critical window with $|x|\in\mathcal{O}(1)$, as well as the crossover critical regimes for both subcritical and supercritical sides, with $|x|\sim L^{X_\text{m}-X_\text{v}}\to\infty$.

\begin{figure}
\centering
\includegraphics[width=1.0\columnwidth]{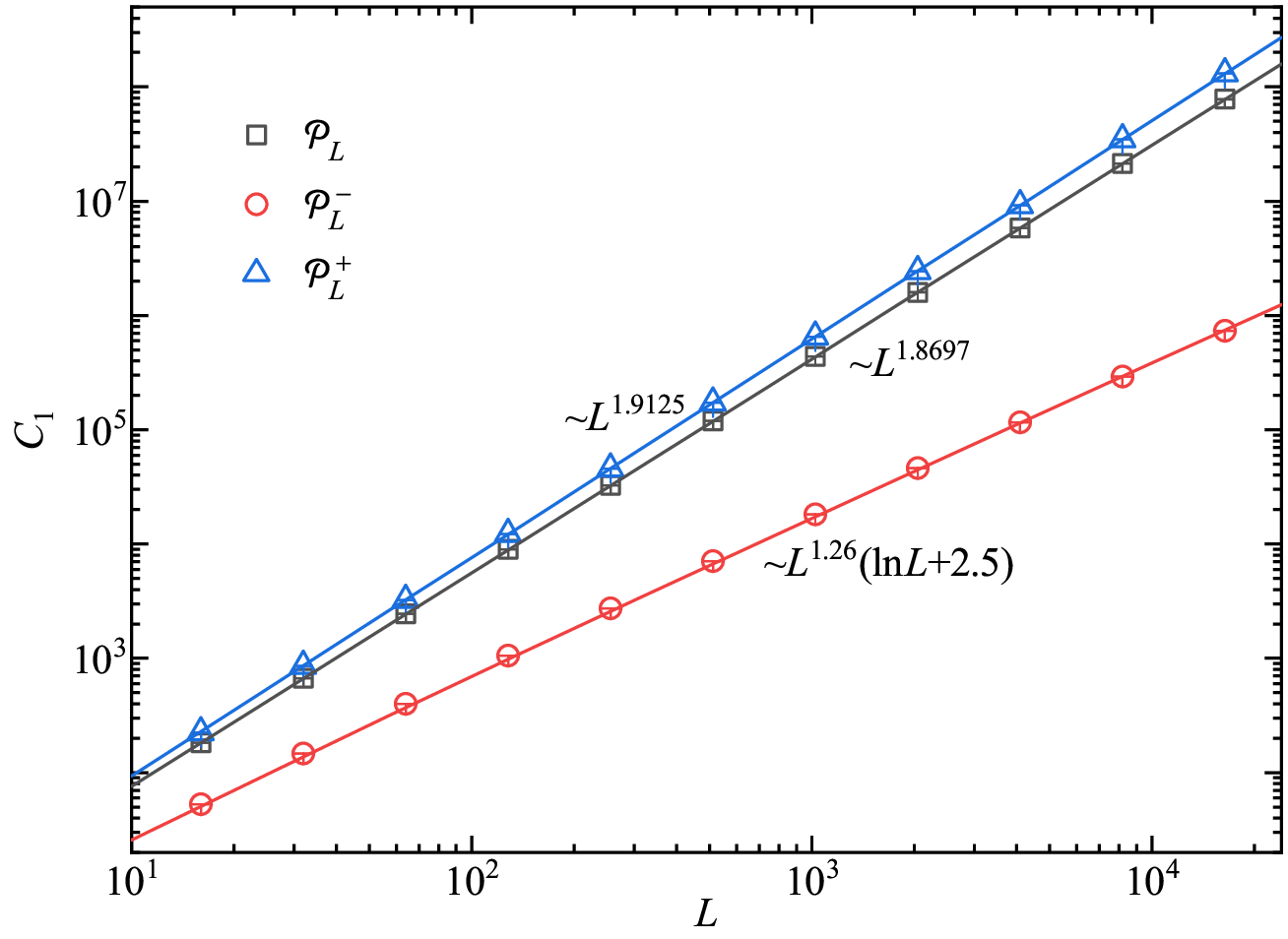}
\caption{(Color online) The size $C_1$ of the largest cluster sampled at $\mathcal{P}_L$ and $\mathcal{P}^\pm_L\equiv\mathcal{P}_L\pm aL^{-\lambda}$ with $\lambda=1$ and $a=1$ for explosive percolation on complete graphs. The fit results suggest the fractal dimension $d_f=1.8697(1)$. From Eqs.~(\ref{eq-dlup}) and (\ref{eq-dllow}), the fractal dimensions at $\mathcal{P}_L^\pm$ are $d_\lambda^+=2-(2-d_f)\lambda\nu\approx1.912$ and $d_\lambda^-=\lambda\nu d_f\approx1.263$, which are consistent with the fit results $d_\lambda^+=1.9125(2)$ and $d_\lambda^-=1.26(2)$ indicated by lines. As the prediction of crossover FSS theory, i.e., Eq.~(\ref{eq-c1l}), there is a multiplicative logarithmic correction for the FSS of $C_1$ at $\mathcal{P}_L^-$.} \label{fg8}
\end{figure}

To further demonstrate the fractal dimensions $d_\lambda^{\pm}$, we present the FSS of $C_1$ sampled at the dynamic bond densities $\mathcal{P}_{L}^\pm \equiv \mathcal{P}_{L} \pm aL^{-\lambda}$ with $\lambda=X_\text{v}=1$ in Fig.~\ref{fg8}, where we set $a=1$ without loss of generality. It can be observed that the size of the largest cluster in the crossover critical regime is well described by Eqs.~(\ref{eq-c1u}) and (\ref{eq-c1l}). By fitting the data extracted at $\mathcal{P}_{L}^-$ to the scaling ansatz Eq.~(\ref{eq-fit2}), and the data extracted at $P_c$ and $\mathcal{P}_{L}^+$ to the scaling ansatz Eq.~(\ref{eq-fit1}), we recover the fractal dimensions $d_f=1.8697(1)$, $d_\lambda^+=1.9125(2)$, and $d_\lambda^-=1.26(2)$.

\subsection{High-dimensional percolation}

\begin{figure}
\centering
\includegraphics[width=\columnwidth]{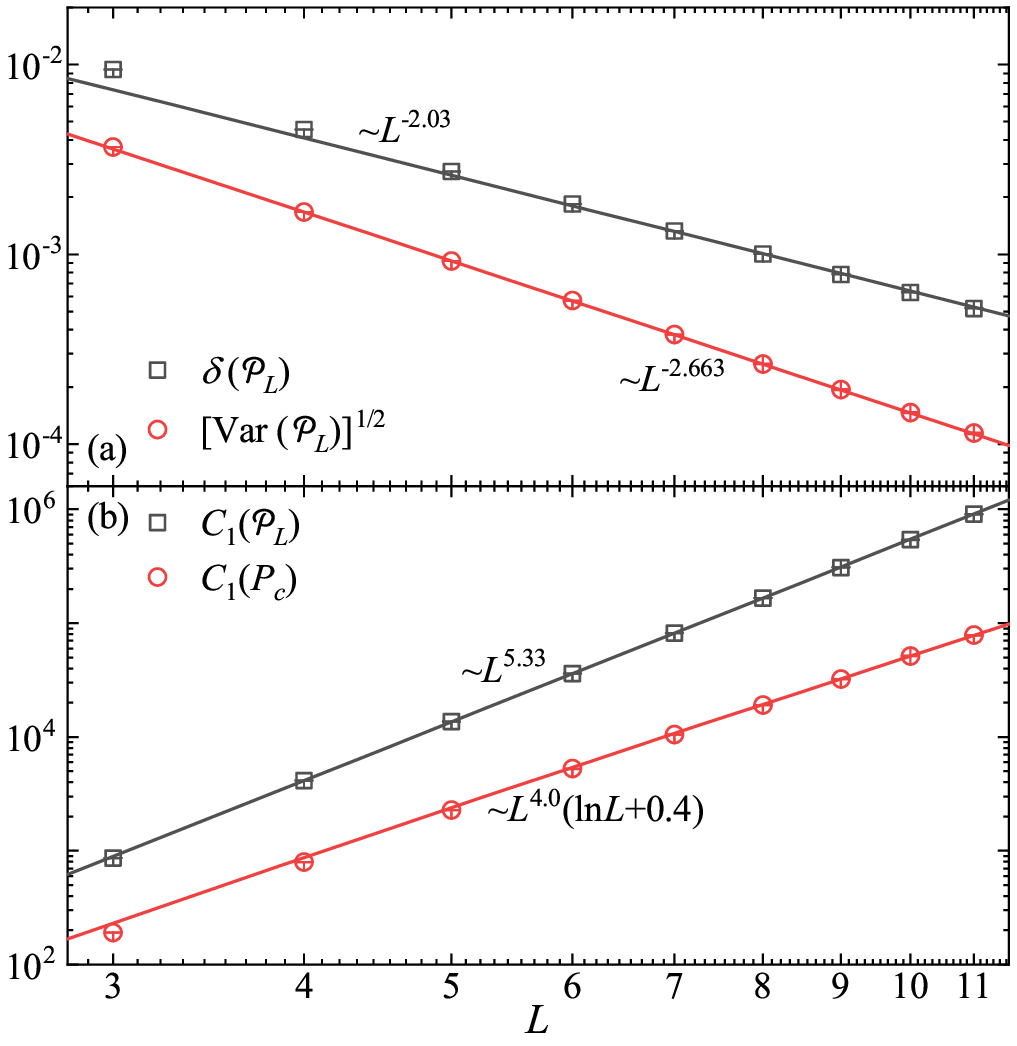}
\caption{(Color online) The finite-size behaviors of the bond percolation on the $8$-dimensional hypercubic lattice with the free boundary condition along one direction and the periodic boundary condition along the remaining directions. (a) The mean deviation $\delta(\mathcal{P}_L)\sim L^{-X_\text{m}}$ of $\mathcal{P}_L$ from $P_c$, and its fluctuation $\sqrt{\text{Var}(\mathcal{P}_L)}\sim L^{-X_\text{v}}$ as functions of the side length $L$. The lines show the fit results $X_\text{m}=2.03(3)$ and $X_\text{v}=2.663(3)$, obtained by fitting the data to the scaling ansatz Eq.~(\ref{eq-fit1}) with $a_2=0$. (b) The FSS of the size $C_1$ of the largest cluster at $\mathcal{P}_L$ and $P_c=0.0677084181$~\cite{Mertens2018}. The lines show the fit results $d_f=5.33(1)$ and $d_f=4.0(1)$, obtained by fitting the data to the scaling ansatz Eqs.~(\ref{eq-fit1}) and (\ref{eq-fit2}) with $a_2=0$, respectively. Note that for $P_c$, a multiplicative logarithmic correction is required to get a better fit of the FSS of $C_1$.} \label{fg9}
\end{figure}

In high-dimensional percolation with $d>d_u=6$, the FSS behaviors are subtle and depend on the boundary condition employed. With periodic boundary conditions, theoretical arguments suggest that the approach of the pseudocritical point to the infinite-system critical point follows the same rate of the vanishing of its fluctuation, scaling as $\sim L^{d/3}$, similar to low dimensions ($d<d_u$)~\cite{Aharony1995}. Indeed, by framing the percolation process as a bond-insertion process, akin to what was described for explosive percolation in the previous section, we can also define exponents $X_\text{m}$ and $X_\text{v}$ for high-dimensional percolation. In this framework, such a scenario can be succinctly described as $X_\text{m}=X_\text{v}=d/3$. However, under free boundary conditions, the mean deviation of $\mathcal{P}_L$ from $P_c$ is governed by a dimension-independent exponent $X_\text{m}=2$, while the exponent $X_\text{v}=d/3$ retains the value observed with periodic boundary conditions \cite{Aharony1995}.

Aharony et al. also proposed that the hyperscaling relation breaks down for dimensions $d>d_u$~\cite{Aharony1984}, and suggested that the spanning cluster at $P_c$ is constrained to a fractal dimension $d_f=4$, which contrasts with the dimension-dependent value observed for $d<d_u$. Coniglio provided a geometric interpretation for this breakdown in the hyperscaling relation~\cite{Coniglio1985}, attributing it to an infinite number of spanning clusters of order $L^{d_f}$, which is proportional to $L^{d-6}$. This interpretation has been conjectured to help understand the FSS of high-dimensional percolation with free boundary conditions at $P_c$. For systems with periodic boundary conditions, it has been conjectured that $d_f=2d/3$ at $P_c$~\cite{Aizenman1997}. However, theoretical discussions suggest that the fractal dimension at the pseudocritical point $P_L$ remains $d_f=2d/3$ for both free and periodic boundary conditions~\cite{Aizenman1997,Kenna2017}. 

We conduct simulations of bond percolation on an $8$-dimensional hypercubic lattice with the free boundary condition along one direction and the periodic boundary condition along the remaining directions. The results are illustrated in Fig.~\ref{fg9} (a) for $\delta(\mathcal{P}_L)$ and $\sqrt{\text{Var}(\mathcal{P}_L)}$. Despite the relatively limited linear system size achieved in the simulation for $d=8$, the algebraic decaying behaviors of $\delta(\mathcal{P}_L)$ and $\sqrt{\text{Var}(\mathcal{P}_L)}$ are discernible from the approximately linear trends observed in the log-log plot in Fig.~\ref{fg9} (a). Furthermore, it is evident that the rate at which $\sqrt{\text{Var}(\mathcal{P}_L)}$ vanishes is notably faster than that of $\delta(\mathcal{P}_L)$. By fitting the data to the scaling ansatz Eq.~(\ref{eq-fit1}) with $a_2=0$, we obtain the exponents $X_\text{m}=2.03(3)$ and $X_\text{v}=2.663(3)$, aligning closely with the values, $X_\text{m}=2$ and $X_\text{v}=8/3\approx2.667$, from theoretical argument~\cite{Aharony1995}.

Furthermore, as shown in Fig.~\ref{fg9} (b), the largest-cluster size demonstrates different fractal dimensions at $\mathcal{P}_L$ and $P_c$. By fitting the data of $C_1(\mathcal{P}_L)$ to Eq.~(\ref{eq-fit1}), a stable fit is obtained as $d_f=5.33(3)$, consistent with the theoretical value $d_f=2d/3$. For $C_1(P_c)$, a multiplicative logarithmic correction is necessary to attain a stable fit. Thus, we fit the data to the scaling ansatz Eq.~(\ref{eq-fit2}). We find the stable fitting result $d_f=4.0(1)$, which is also consistent with the theoretical value $d_f=4$.

The relationship of $X_\text{m}<X_\text{v}$, which is different from explosive percolation, means that the dynamic pseudocritical point $\mathcal{P}_L$ fluctuates in a narrow window $\mathcal{O}(L^{-X_\text{v}})$, while its mean is far away from the infinite-system critical point $P_c$, scaling as $\sim L^{-X_\text{m}}$. As shown in Fig.~\ref{fg11}, by plotting $C_1/L^{d_f}$ and $\chi/L^{2d_f-d}$ as functions of $(P-\mathcal{P}_L)L^{X_\text{v}}$, the data from different system sizes collapse on top of each other well. Hence, we argue that the FSS theory is also well applied around $\mathcal{P}_L$ for high-dimensional percolation, with the correlation-length exponent $\nu=1/X_\text{v}$, and the FSS extracted at $P_c$ corresponds to the lambda-scaling in the crossover critical regime respected to $\mathcal{P}_L$.

Due to free boundary conditions, the dynamic pseudocritical point is anticipated and numerically confirmed to occur at $\mathcal{P}_L>P_c$, thus, the lambda-scaling of fractal dimension at $P_c$ is predicted by Eq.~(\ref{eq-dllow}), which immediately recovers $d_\lambda=\lambda\nu d_f=4$ by letting $d_f=2d/3$, $\nu=1/X_\text{v}=3/d$, and $\lambda=X_\text{m}=2$. Moreover, the discussion on Eq.~(\ref{eq-c1l}) also explains the origin of the multiplicative logarithmic correction in the FSS of $C_1$ at $P_c$ as shown in Fig.~\ref{fg9} (b). Furthermore, Eq.~(\ref{eq-ncl}) also explains that the number of clusters of size $\mathcal{O}(L^4)$ is divergent as $\sim L^{d-6}$ at $P_c$.

\begin{figure}
\centering
\includegraphics[width=1.0\columnwidth]{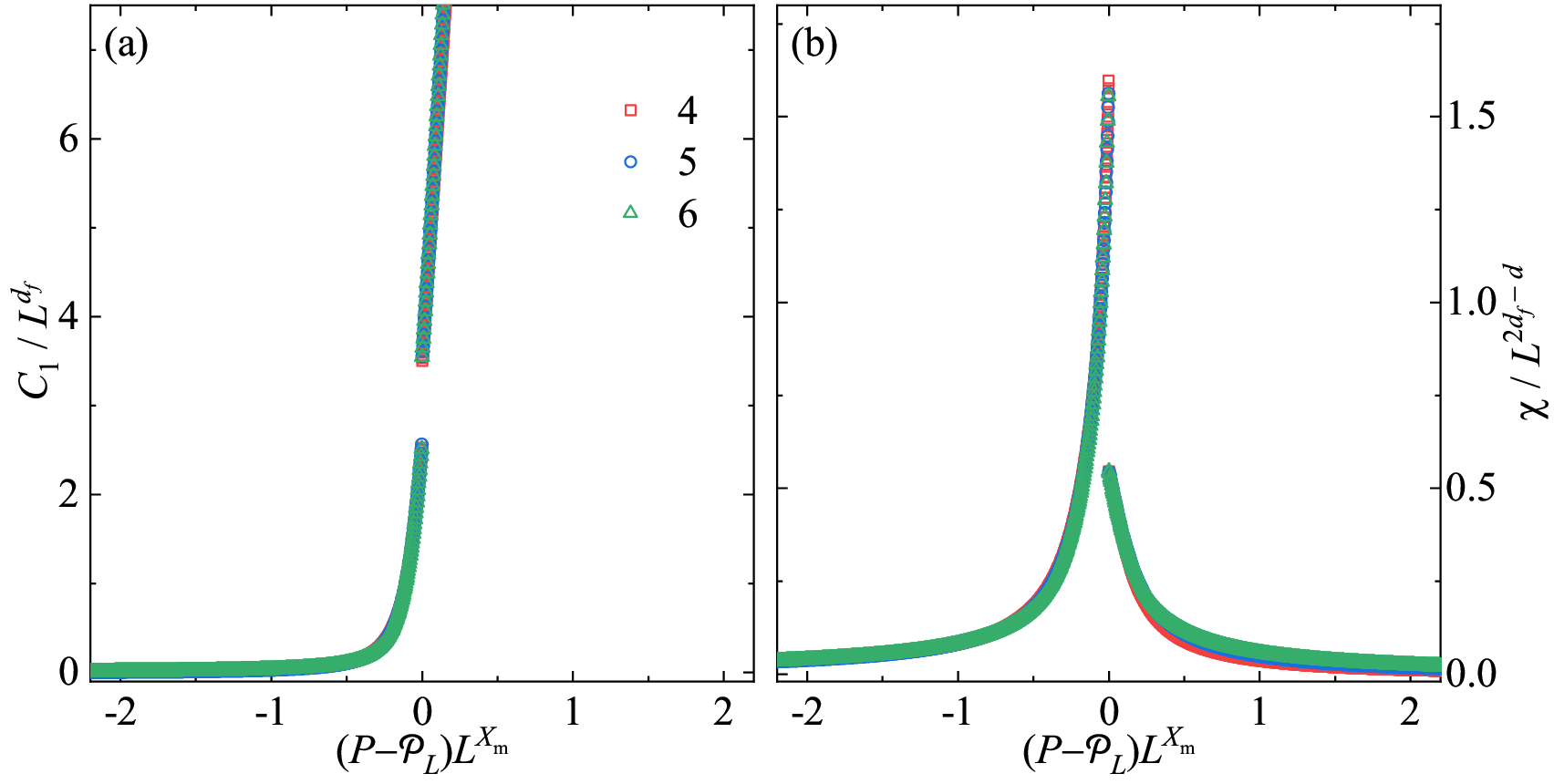}
\caption{(Color online) The data collapse near the dynamic pseudocritical point $\mathcal{P}_L$ for bond percolation on an $8$-dimensional hypercubic lattice with the free boundary condition along one direction and the periodic boundary condition along the remaining directions. (a) The plots of $C_1/L^{d_f}$ versus $(P-\mathcal{P}_L)L^{1/\nu}$ with $1/\nu=X_\text{v}=d/3$ and $d_f=2d/3$ for different $L$. (b) The plots of $\chi/L^{2d_f-2}$ versus $(P-\mathcal{P}_L)L^{1/\nu}$ with $1/\nu=X_\text{v}=d/3$ and $d_f=2d/3$ for different $L$. The discontinuity at $\mathcal{P}_L$ arises from its event-based definition.} \label{fg10}
\end{figure}

\subsection{The classification of percolation systems based on their finite-size behaviors}

\begin{table*}
\caption{Critical exponents for the asymptotic behaviors of the dynamic pseudocritical point $\mathcal{P}_L$. In the simulation, the percolation process is realized by inserting occupied bonds one by one. During the bond-insertion process, two special events are utilized to identify the dynamic pseudocritical point, i.e., the maximum one-step incremental size of the largest cluster $\Delta$ and the maximum of the susceptibility $\chi$. The exponents are obtained by fitting the data to the scaling ansatz Eq.~(\ref{eq-fit1}) with $a_2=0$. For $2$-dimensional percolation, the theoretical values are $X_\text{m}=X_\text{v}=1/\nu=3/4$. For high-dimensional percolation with periodic boundary conditions (PBC), $X_\text{m}=X_\text{v}=1/\nu=d/3$, while with free boundary conditions (FBC), $X_\text{m}=2$ and $X_\text{v}=1/\nu=d/3$. As mentioned in the main text, the finite-size behaviors on complete graphs are defined with respect to the side length $L\equiv\sqrt{V}$ with $V$ the total number of sites, so that, the standard bond percolation on complete graphs has $X_\text{m}=X_\text{v}=1/\nu=2/3$.}   \label{tb-pl}
\begin{ruledtabular}
\begin{tabular}{cccccccc}
Method & Exponent  & \multicolumn{4}{c}{Bond percolation on $d$-dimensional hypercubic lattices}  & \multicolumn{2}{c}{Complete graphs}  \\
       &    & $d=2$ PBC & $d=8$ PBC  & $d=8$ FBC & $d=9$ FBC  &  Bond percolation  &  Explosive percolation \\
\hline
Maximum $\Delta$  &  $X_\text{m}$     & 0.750(1)  & 2.66(1)~~  & 2.03(3)~~  & 2.01(5)  & 0.666(6)~~  & 1.480(4)     \\
                  &  $X_\text{v}$  & 0.748(2)  & 2.669(4)   & 2.663(3)   & 2.99(6)  & 0.6670(6)   & 1.006(6)     \\
\hline
Maximum $\chi$    &  $X_\text{m}$     & 0.750(1)  & 2.65(1)~~  & 2.04(4)~~  & 2.01(3)  & 0.664(4)~~  & 1.4(1)~~~~       \\
                  &  $X_\text{v}$  & 0.746(1)  & 2.67(1)~~  & 2.65(1)~~  & 2.98(4)  & 0.667(8)~~  & 1.000(2)   \\
\end{tabular}
\end{ruledtabular}
\end{table*}

Explosive percolation and high-dimensional percolation with free boundary conditions demonstrate two special FSS for $1/\nu= X_\text{m}>X_\text{v}$ and $X_\text{m}<X_\text{v}=1/\nu$, respectively. Together with the common scenario $X_\text{m}=X_\text{v}=1/\nu$, in principle, there exists three types of percolation systems based on this FSS perspective.

\begin{figure}
\centering
\includegraphics[width=1.0\columnwidth]{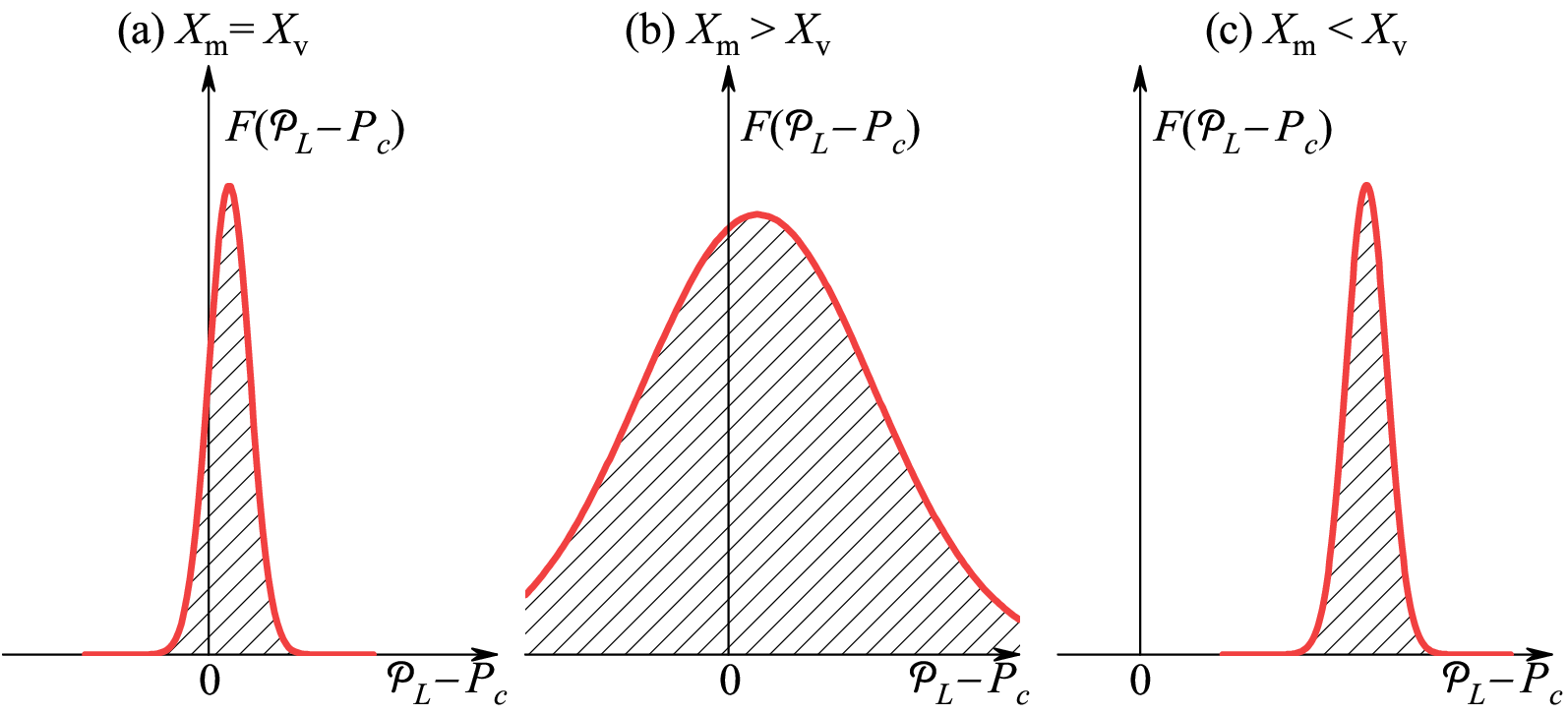}
\caption{(Color online) A sketch of the three possible distributions of the dynamic pseudocritical point $\mathcal{P}_L$, demonstrated by the distribution $F(\mathcal{P}_L-P_c)$, with $P_c$ the infinite-system critical point. The asymptotic behaviors of $\mathcal{P}_L$ are described by Eqs.~(\ref{eq-plm}) and (\ref{eq-plsd}), i.e., the mean deviation of $\mathcal{P}_L$ from $P_c$, $\delta(\mathcal{P}_L) \sim L^{-X_\text{m}}$, and its variance, $\text{Var}(\mathcal{P}_L)\sim L^{-2X_\text{v}}$. (a) $X_\text{m}=X_\text{v}=1/\nu$. Both $\delta(\mathcal{P}_L)$ and $\sqrt{\text{Var}(\mathcal{P}_L)}$ scale as $\sim L^{-1/\nu}$, so that the FSS ansatz Eq.~(\ref{eq-otl}) can be equally well applied to $\mathcal{P}_L$, $P_c$, and $P_L$. (b) $X_\text{v}<X_\text{m}=1/\nu$. Although the mean deviation of $\mathcal{P}_L$ from $P_c$ is governed by the correlation-length exponent $\nu$, the fluctuation of $\mathcal{P}_L$ is so large that, in any single realization, $\mathcal{P}_L$ can be much larger than, close to, or much smaller than the value of $P_c$ (or $P_L$). As a consequence, the FSS behaviors at $P_c$ and $P_L$ can be sophisticated, due to the mixing effects of those in a wide range near $\mathcal{P}_L$. (c) $X_\text{m}<X_\text{v}=1/\nu$. The fluctuation of $\mathcal{P}_L$ is governed by the correlation-length exponent $\nu$, and, thus, the FSS behaviors at the dynamic and static pseudocritical points, $\mathcal{P}_L$ and $P_L$, take the same form. However, in comparison with the fluctuation of $\mathcal{P}_L$, the distance from $P_c$ to $P_L$ is too far, i.e., $L^{-X_\text{m}} \gg L^{-1/\nu}$, and the FSS behaviors at $P_c$ correspond to those dynamic ones outside the finite-size critical window, approaching $\mathcal{P}_L$ in a speed of $L^{-X_\text{m}}$.}
\label{fg11}
\end{figure}

For $X_\text{m}=X_\text{v}=1/\nu$, both the mean deviation of $\mathcal{P}_L$ from $P_c$ and its fluctuation are governed by the correlation-length exponent $\nu$. This implies that, in any given realization, the separation between $P_c$ and $\mathcal{P}_L$ cannot exceed the finite-size critical window $\mathcal{O}(L^{-1/\nu})$. Consequently, the FSS ansatz given by Eq.~(\ref{eq-otl}) can be equally well applied to $P_c$, $P_L$, and $\mathcal{P}_L$. This scenario is visually depicted in Fig.~\ref{fg11} (a) and applies to most cases, including the standard percolation in dimensions $d<d_u$, in dimensions $d>d_u$ with periodic boundary conditions, and on complete graphs, as outlined in Tab.~\ref{tb-pl}.

The second scenario of finite-size behaviors occurs in explosive percolation, where $X_\text{m}=1/\nu$ and $X_\text{v}<1/\nu$. In this scenario, while the mean deviation $\delta(\mathcal{P}_L)$ vanishes rapidly as $\mathcal{O}(L^{-1/\nu})$, the fluctuation $\sqrt{\text{Var}(\mathcal{P}_L)}$ is substantial and decreases at a slower rate, as $\sim L^{-X_\text{v}}$, where $L^{-X_\text{v}} \gg L^{-1/\nu}$. Consequently, the finite-size scaling behaviors at a fixed bond density, such as $P_c$ or $P_L$, can be interpreted as a stochastic mixture of those in a considerably wide range around $\mathcal{P}_L$, see a sketch in Fig.~\ref{fg11} (b).

The third scenario of finite-size behaviors applies to high-dimensional percolation with free boundary conditions, where $X_\text{v}=1/\nu =d/3$ and $X_\text{m}=2<1/\nu$. In other words, the fluctuation of the dynamic pseudocritical point $\mathcal{P}_L$ is governed by the correlation-length exponent $\nu$, but the infinite-system critical point $P_c$ is approached at a much slower rate as $\sim L^{-X_\text{m}}\gg L^{-1/\nu}$, as depicted in Fig.~\ref{fg1} (c). In this scenario, the FSS behaviors at the static pseudocritical point $P_L$ mirror those at the dynamic pseudocritical point $\mathcal{P}_L$. However, the FSS behaviors at the infinite-system critical point $P_c$ correspond to those outside the finite-size critical window centred on $\mathcal{P}_L$.

Moreover, the above classification of the percolation systems is independent of the definition of the dynamic pseudocritical point $\mathcal{P}_L$. One can alternatively define the dynamic pseudocritical point as the maximum point of susceptibility $\chi$ in a single realization. In Tab.~\ref{tb-pl}, we also summarize the fit results of $X_\text{m}$ and $X_\text{v}$ for such a definition of the dynamic pseudocritical point. It is cleat that the exponents $X_\text{m}$ and $X_\text{v}$ are consistent regardless of the definition of the dynamic pseudocritical point $\mathcal{P}_L$. This observation suggests that the asymptotic behavior of the dynamic pseudocritical point reveals fundamentally different FSS behaviors and might serve as a crucial starting point for comprehending various sophisticated FSS behaviors observed in the conventional ensemble.

\section{Conclusion}  \label{sec-dis}

In this paper, we undertake on a systematic investigation of FSS behaviors within a crossover critical regime, which refers to the region $(t\to0,x\to\infty)$ in the standard FSS ansatz $Q(t,L)=L^{Y_Q}\tilde{Q}(x\equiv tL^{1/\nu})$. By ensuring consistency between FSS and thermodynamic scaling, we establish the asymptotic behavior of the universal function $\tilde{Q}(x\to\infty)$, validated by simulation results from percolation systems. Consequently, we comprehensively depict the crossover finite-size behavior from FSS to thermodynamic scaling and establish relationships between the critical exponents of infinite-system critical and FSS behaviors. These findings, termed crossover FSS theory, lead to a significant corollary — the finite-size lambda-scaling concept. This concept reveals that FSS phenomena can be observed across a broader range of positions as $t\to0$ for $L\to\infty$; when $t$ exceeds the finite-size critical window, the exponents describing FSS behaviors vary accordingly.

As applications, the crossover FSS theory can quantitatively elucidate the different FSS behaviors extracted at pseudocritical and infinite-system critical points observed in various percolation systems, including explosive percolation and high-dimensional percolation. We demonstrate that accurate FSS can consistently be extracted at the dynamic pseudocritical point $\mathcal{P}_L$. However, in the conventional ensemble (at $P_c$ or $P_L$), FSS might exhibit complexity depending on the distribution of $\mathcal{P}_L$, which can be quantitatively understood using lambda-scaling. From these observations, percolation systems with various FSS fall into three classifications.

Moreover, it is pertinent to highlight several intriguing and noteworthy questions. Firstly, our findings underscore the necessity of discerning between universal properties observed at critical points defined differently when applying the FSS theory to extract such properties for systems undergoing continuous phase transitions. Secondly, it is noteworthy that the FSS theory finds optimal application in the vicinity of the dynamic pseudocritical point rather than the infinite-system critical point. Thirdly, the crossover FSS theory, being a universal framework for phase transitions, extends beyond percolation transitions. Hence, our findings offer crucial insights into bridging critical behaviors across different ensembles for diverse phase transitions.

\section*{Acknowledgements}

The research was supported by the National Natural Science Foundation of China under Grant No.~12275263, the Innovation Program for Quantum Science and Technology under Grant No.~2021ZD0301900, and Natural Science Foundation of Fujian province of China under Grant No.~2023J02032. The research of M.L. was also supported by the Fundamental Research Funds for the Central Universities (No.~JZ2023HGTB0220).

\bibliography{ref}

\end{document}